\pdfoutput=1
\documentclass[acmtog]{acmart}

\setcitestyle{square}

\usepackage{amsmath,amsthm,amssymb}
\usepackage{graphicx}
\usepackage{wrapfig}
\usepackage{color}
\usepackage{xspace}
\usepackage{overpic}
\usepackage{subfig}
\usepackage{wrapfig}
\usepackage{enumitem}
\usepackage{tabularx}
\usepackage{caption}
\usepackage{soul}

\definecolor{turquoise}{cmyk}{0.65,0,0.1,0.1}
\definecolor{purple}{rgb}{0.65,0,0.65}
\definecolor{dark_green}{rgb}{0, 0.5, 0}
\definecolor{blue}{rgb}{0, 0, 1.0}
\definecolor{orange}{rgb}{0.8, 0.6, 0.2}
\definecolor{red}{rgb}{0.8, 0.2, 0.2}
\definecolor{brown}{rgb}{0.5, 0.16, 0.16}

\newcommand{\revised}[1]{{\leavevmode\color{blue}#1}}

\begin{document}

\title{GRAINS: Generative Recursive Autoencoders for INdoor Scenes}

\author{Manyi Li}
\affiliation{%
	\institution{Shandong University and Simon Fraser University}
}
\author{Akshay Gadi Patil}
\affiliation{%
	\institution{Simon Fraser University}
}
\author{Kai Xu}
\authornote{Corresponding author: kevin.kai.xu@gmail.com}
\affiliation{%
	\institution{National University of Defense Technology and AICFVE Beijing Film Academy}
}
\author{Siddhartha Chaudhuri}
\affiliation{%
	\institution{Adobe Research and IIT Bombay}
}

\author{Owais Khan}
\affiliation{%
	\institution{IIT Bombay}
}
\author{Ariel Shamir}
\affiliation{%
	\institution{The Interdisciplinary Center}
}
\author{Changhe Tu}
\affiliation{%
	\institution{Shandong University}
}
\author{Baoquan Chen}
\affiliation{%
	\institution{Peking University}
}
\author{Daniel Cohen-Or}
\affiliation{%
	\institution{Tel Aviv University}
}
\author{Hao Zhang}
\affiliation{%
	\institution{Simon Fraser University}
}


\renewcommand\shortauthors{M. Li et al}

\begin{abstract}
We present a {\em generative neural network\/} which enables us to generate plausible 3D indoor scenes
in large quantities and varieties, easily and highly efficiently. Our key observation is that indoor scene structures are
inherently {\em hierarchical\/}. Hence, our network is not convolutional; it is a {\em recursive\/} neural network
or RvNN. Using a dataset of annotated scene hierarchies, we train a {\em variational recursive autoencoder\/}, or
RvNN-VAE, which performs scene object grouping during its encoding phase and scene generation during
decoding. Specifically, a set of encoders are recursively applied to group 3D objects based on support, surround,
and co-occurrence relations in a scene, encoding information about objects' spatial properties, {\em semantics\/},
and their {\em relative\/} positioning with respect to other objects in the hierarchy. By training a variational autoencoder (VAE),
the resulting fixed-length codes roughly follow a Gaussian distribution. A novel 3D scene can be generated
hierarchically by the decoder from a randomly sampled code from the learned distribution.
We coin our method GRAINS, for Generative Recursive Autoencoders for INdoor Scenes. We demonstrate the
capability of GRAINS to generate plausible and diverse 3D indoor scenes and compare with existing methods for 3D scene
synthesis. We show applications of GRAINS including 3D scene modeling from 2D layouts, scene editing,
and semantic scene segmentation via PointNet whose performance is boosted by the large quantity
and variety of 3D scenes generated by our method.

\end{abstract}

%
%

\ccsdesc[500]{Computing methodologies~Computer graphics}
\ccsdesc[300]{Computing methodologies~Shape analysis}


\keywords{3D indoor scene generation, recursive neural network, variational autoencoder}



\begin{teaserfigure}
	\includegraphics[width=\textwidth]{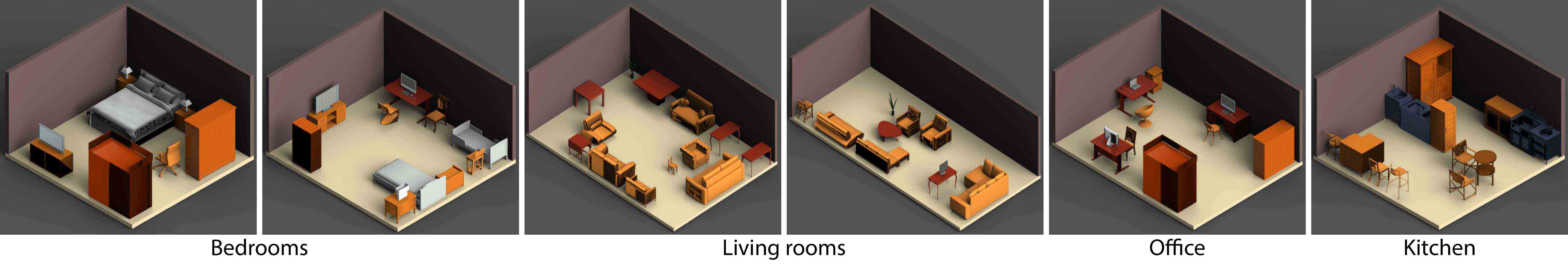}
	\caption{We present a generative recursive neural network (RvNN) based on a variational autoencoder (VAE) to learn hierarchical scene structures, enabling us to easily generate plausible 3D indoor scenes in large quantities and varieties (see scenes of kitchen, bedroom, office, and living room generated). Using the trained RvNN-VAE, a novel 3D scene can be generated from a random vector drawn from a Gaussian distribution in a fraction of a second.}
	\label{fig:teaser}
\end{teaserfigure}

\maketitle




\section{Introduction}
\label{sec:intro}

With the resurgence of virtual and augmented reality (VR/AR), robotics, surveillance, and smart homes,
there has been an increasing demand for virtual models of 3D indoor
environments. 
At the same time, modern approaches to solving many scene analysis and modeling problems have
been data-driven, 
resorting to machine learning. 
More training data, in the form of structured and annotated 3D indoor scenes, can directly boost the
performance of learning-based methods. All in all, the era of ``big data'' for 3D indoor scenes is
seemingly upon us.
Recent works have shown that generative neural networks can be trained to synthesize
images, speech, and 3D shapes. 
An obvious
question is whether similar success can be achieved with 3D indoor scenes, enabling us to
easily generate a large number of 3D scenes that are {\em realistic\/} and {\em diverse\/}.

The basic building blocks of an indoor scene are 3D objects whose semantic or functional composition and
spatial arrangements follow clear patterns, but still exhibit rich structural variations even within the same
scene category (e.g., think of the layout varieties of kitchens or bedrooms). Highly structured models, including
indoor scenes and many man-made objects, could be represented as a volume or using multi-view images and
undergo conventional convolutionary operations. However, such operations are oblivious to the underlying
structures in the data which often play an essential role in scene understanding. This may explain in part why
deep convolutional neural networks (CNNs), which have been so successful in processing natural images, have
not been widely adopted for the analysis or synthesis of 3D indoor scenes.

\begin{figure*}[!t]
        \includegraphics[width=0.9\linewidth]{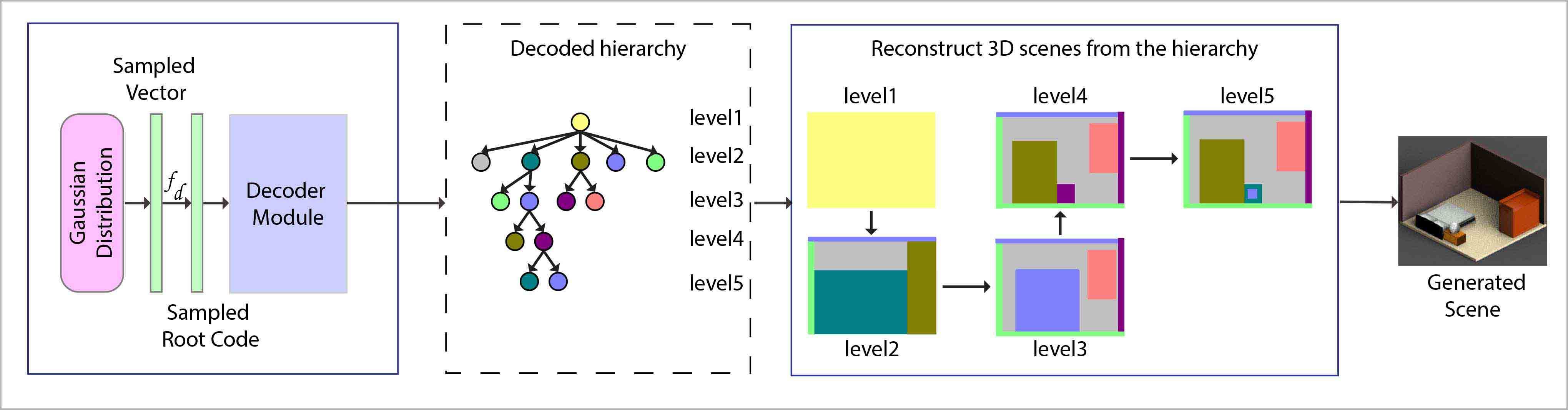}
        \caption{Overall pipeline of our scene generation. The decoder of our trained RvNN-VAE turns a randomly sampled code from the learned distribution into a plausible indoor scene hierarchy composed of OBBs with semantic labels. The labeled OBBs are used to retrieve 3D objects to form the final 3D scene.}
        \label{fig:pipeline}
\end{figure*}

In this paper, we present a 
{\em generative neural network\/} which enables us to easily generate plausible 3D indoor scenes in large
quantities and varieties; see Figure~\ref{fig:teaser}. Using our approach, a novel 3D scene can be generated
from a random vector drawn from a Gaussian in a fraction of a second,
following the pipeline shown in Figure~\ref{fig:pipeline}. Our key observation is that indoor
scene structures are inherently {\em hierarchical\/}. Hence, our network is not convolutional;
it is a {\em recursive\/} neural network~\cite{socher2011} or RvNN\footnote{RvNNs are not to be confused with
recurrent neural networks or RNNs.}.

\begin{figure}[!t]
	\includegraphics[width=\linewidth]{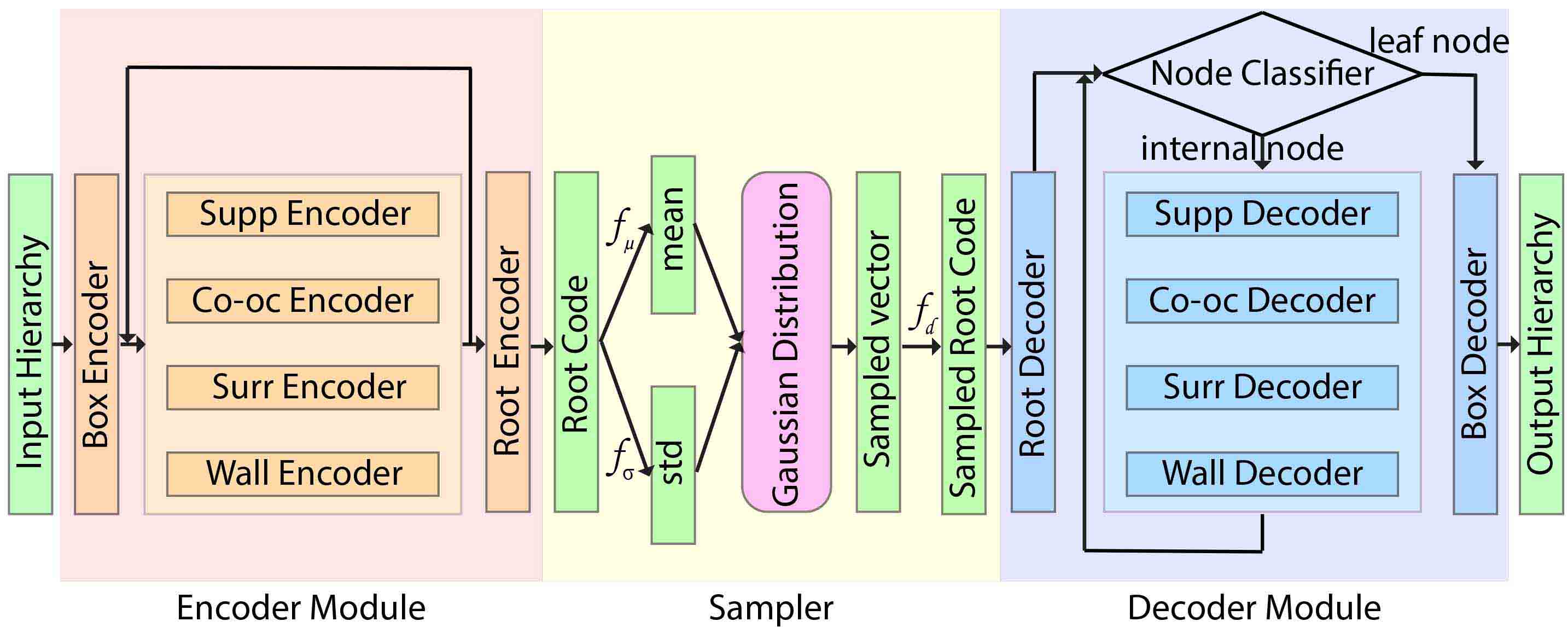}
	\caption{Architecture of our RvNN-VAE, which is trained to learn a generative model of indoor scene structures.
Input to the network is a scene hierarchy composed of labeled OBBs enclosing 3D scene objects.
The boxes are recursively grouped and codified by a set of encoders, resulting in a root code.
The root code is approximated by a Gaussian distribution, from which a random vector is drawn and fed to the
decoder. The recursive decoder produces an output hierarchy to minimize a reconstruction+VAE loss.}
	\label{fig:network_struct}
\end{figure}

Using a dataset of annotated scene hierarchies, we train a {\em variational recursive autoencoder\/}, or RvNN-VAE,
which performs scene object grouping during its encoding phase and scene generation during decoding, as shown
in Figure~\ref{fig:network_struct}. Specifically, a set of encoders are recursively applied to group 3D objects in a
scene, bottom up, and encode information about the objects and their relations, where the resulting
fixed-length codes roughly follow a Gaussian distribution. A new scene can then be generated top-down, i.e.,
hierarchically, by decoding from a randomly generated code.

Our approach is inspired by the recent work of Li et al.~\shortcite{li2017}, coined GRASS, which develops
a generative recursive autoencoder for learning shape structures. Specifically, they model the part structure
of a 3D object using a hierarchical organization of assembly and symmetry groupings over the object parts
and train an RvNN built on autoencoders to learn hierarchical grouping rules. Like GRASS, our neural network
for indoor scenes, which we call {\em GRAINS\/}, is also an RvNN-based generative autoencoder. But the data
representation, network architecture, and training mechanism all have to be altered in fundamental ways
to meet the unique challenges arising from learning indoor scene structures and object-object relations.

On a superficial level, one may regard parts in a 3D object as the same as objects in a 3D scene. While
part-part and object-object relations, for objects and scenes respectively, are both highly structured,
there are several critical distinctions between them:
\begin{itemize}
\item Predominantly, the constituent parts of a 3D object are strictly governed by connectivity and
symmetry~\cite{wang2011}. In contrast, relations between objects in a scene are much {\em looser\/}.
The objects are almost never physically connected. As well, due to perturbations arising from
human usage, symmetries (e.g., the rotational symmetry of dining chairs around a round table) are
often not strictly adhered to.
\item Object-object relations in a scene are often more about {\em semantics\/} or {\em functionality\/}
than geometric proximity or fit, with the latter being more prominent for part-part relations in 3D object
geometry. For example, a TV and a sofa are related since they together serve the function of ``watching
TV", but the two objects can be far apart in a scene. Unlike symmetry and proximity, relations of this kind
are much more difficult, if not impossible, to infer based purely on geometric analysis or by analyzing a given
scene alone.
\item Despite the rich intra-class structural variations that exist with man-made objects, e.g., the variety of
chairs or airplanes, these objects can still be well aligned, at least globally. Such an alignment offers predictability
in the positioning of object parts, which often plays an essential role in facilitating the learning of
models or templates for man-made shapes~\cite{kim2013,huang2015,li2017}. In contrast, it is
significantly more difficult to align scene layouts, even if they are all of bedrooms
or living rooms.
\end{itemize}

To address the new challenges as a result of these differences, GRAINS differs from GRASS in several
key aspects. First, we construct our RvNN for indoor scenes using three grouping operations (with their
corresponding encoders and decoders as shown in Figure~\ref{fig:network_struct}):
{\em support\/} (e.g., desk supporting a computer), {\em surround\/}
(e.g., nightstands surrounding a bed or dining chairs surrounding a table), and {\em co-occurrence\/}
(e.g., between sofa and coffee table or between computer, keyboard, and mouse). In contrast, the GRASS
RvNN for 3D shapes is defined by two grouping operations: symmetry and connectivity.
Generally speaking, the term ``object co-occurrence'' would encompass both support
and surround relations. However, in our paper, co-occurrences is a ``catch-all'' entity that covers all the
geometrically loose and functionality- or action-oriented object-object relations that do {\em not\/} reflect
physical support or surround.

Second, a proper grouping of 3D objects in a scene has to account for object co-occurrences, which are inherently
tied to the {\em semantics\/} of the objects. Indeed, co-occurrence hinges mainly on {\em what\/} the objects are,
e.g., a computer monitor is almost always associated with a keyboard. Thus, unlike GRASS, our scene RvNN must
encode and decode both {\em numerical\/} (i.e., object positioning and spatial relations) and
{\em categorical\/} (i.e., object semantics) data. To this end, we use labeled oriented bounding boxes (OBBs)
to represent scene objects, whose semantic labels are recorded by one-hot vectors.

%

Finally, GRASS encodes {\em absolute\/} part positions, since the global alignment between 3D objects belonging to
the same category leads to predictability of their part positions. However, in the absence of any sensible global scene
alignment over our scene dataset, GRAINS must resort to {\em relative\/} object positioning to reveal
the predictability of object placements in indoor scenes. In our work, we encode the relative position of an OBB
based on offset values from the boundaries of a reference box, as well as alignment and orientation attributes
relative to this reference. Under this setting, room walls serve as the initial reference objects (much
like the ground plane for 3D objects) to set up the scene hierarchies. Subsequent reference boxes are determined 
on-the-fly for nested substructures. 


Figure~\ref{fig:network_struct} shows the high-level architecture of our RvNN-VAE and Figure~\ref{fig:pipeline} 
outlines the process of scene generation. We demonstrate that GRAINS enables us to generate a large number of
plausible and diverse 3D indoor scenes, easily and highly efficiently. Plausibility tests are conducted via perceptual studies
and comparisons are made to existing 3D scene synthesis methods. Various network design choices, e.g., semantic 
labels and relative positioning, are validated by experiments. Finally, we show applications of GRAINS including 3D 
scene modeling from 2D layouts, scene manipulation via hierarchy editing, and semantic scene segmentation via 
PointNet~\cite{qi2017pointnet} whose performance is clearly boosted by the large quantity and variety of 3D scenes 
generated by GRAINS.


\section{Related work}
\label{sec:related}

In recent years, much success has been achieved on developing deep neural networks, in particular convolutional
neural networks, for pattern recognition and discriminative analysis of visual data. 
Some recent works have also shown that generative neural networks can be trained to synthesize images~\cite{oord2016_rnn},
speech~\cite{oord2016_wavenet}, and 3D shapes~\cite{wu2016}. The work we present is an attempt at designing
a generative neural network for 3D indoor scenes. As such, we mainly cover related works on modeling
and synthesis of 3D shapes and 3D scenes.

\paragraph{Indoor scene synthesis.}
The modeling of indoor environments is an important aspect of 3D content creation.
The increasing demand of 3D indoor scenes from AR/VR, movies, robotics, etc, calls for effective ways
of automatic synthesis algorithms.
Existing approaches to scene synthesis mostly focus on probabilistic modeling
of object occurrence and placement.
The technique of Fisher et al.~\shortcite{fisher2012} learns two
probabilistic models for modeling sub-scenes (e.g. a table-top scene): (1) object occurrence,
indicating which objects should be placed in the scene, and
(2) layout optimization, indicating where to place the objects.
Given an example scene, new variants can be synthesized based on the learned priors.

Graphical models are recently utilized to model global room layout, e.g.,~\cite{kermani2016learning}.
To ease the creation of guiding examples, Xu et al.~\shortcite{xu2013sketch2scene}
propose modeling 3D indoor scenes from 2D sketches, by leveraging
a database of 3D scenes. Their method jointly optimizes for
sketch-guided co-retrieval and co-placement of scene objects.
Similar method is also used to synthesize indoor scenes from videos or RGB-D images~\cite{chen2014,kermani2016learning}.
Some other works perform layout enhancement of a user-provided initial layout~\cite{yu2011,merrell2011}.
Fu et al.~\shortcite{fu2017} show how to populate a given room with objects
with plausible layout.
A recent work of Ma et al. ~\shortcite{ma2018language} uses language as a prior to guide
the scene synthesis process. Scene edits are performed by first parsing a natural language command from the user and transforming it into a semantic scene graph that is used to retrieve corresponding sub-scenes from a database. This retrieved sub-scene is then augmented by incorporating other objects that may be implied by the scene context. A new 3D scene is synthesized by aligning the augmented sub-scene with the current scene.

Human activity is another strong prior for modeling scene layout.
Based on an activity model trained from an annotated database of scenes and 3D objects,
Fisher et al.~\shortcite{fisher2015} synthesize scenes to fit incomplete 3D scans of real scenes.
Ma et al.~\shortcite{ma2016} introduce action-driven evolution of 3D indoor
scenes, by simulating how scenes are altered by human activities.
The action model is learned with annotated photos, from which
a sequence of actions is sampled to progressively evolve an initial clean 3D scene.
Recently, human activity is integrated with And-Or graphs, forming a
probabilistic grammar model for room layout synthesis~\cite{qi2018human}.

In a concurrently developed work, Wang et al.~\shortcite{wang2018} learn deep convolutional priors for indoor scene synthesis. Their method encodes scene composition and layout using multi-channel top-view depth images. A deep convolutional neural network is then trained with these image channels to output object placement priors as a 2D distribution map. Scene synthesis is performed via a sequential placement of objects, guided by the learned priors. The key differences between their and our scene synthesis frameworks include: 1) GRAINS produces 3D indoor scenes with object-on-furniture support and is not limited to just floor-supported furniture layouts; 
2) we model and learn scene {\em hierarchies\/} rather than a flat object layout; 3) we adopt a structural representation of indoor scenes which explicitly encode spatial relations and contextual information.

\paragraph{Generative 3D modeling.}
We focus on generative models for creating discrete variations of 3D models at the part
level. Earlier models are either based on hand-crafted shape grammars~\cite{Mueller2006},
or learned from a single or a few training examples~\cite{Bokeloh2010,Talton2012}.
More recent models are usually constructed from shape correlation across of a larger set of 3D models,
e.g., feature point correspondence~\cite{kim2013} or part correspondence~\cite{xu2012}.
Based on part correspondence, some works explored the learning of part-based Bayesian networks~\cite{Chaudhuri2011,Kalogerakis2012}
or deformable templates~\cite{Fish2014,huang2015}, to encode both continuous and discrete variations.
Such models have been successfully extended to scene modeling~\cite{fisher2012,fu2017}.

\paragraph{Deep generative models for 3D modeling.}
Deep generative networks including Variational Auto-Encoders (VAE)~\cite{Kingma2013} and generative adversarial nets (GAN)~\cite{Goodfellow2014} have enabled effective modeling of high-dimensional
data with a low-dimensional latent space. New samples can be drawn from the latent space and
mapped back to the original data space with high quality. Limited by the available vector
representations, however deep generative models for 3D shapes have thus far been focusing on
the generation of objects or scenes in volumetric representation~\cite{wu2015,girdhar2016,wu2016,song2016ssc}.
A major drawback of such models is the generated volumes are structure-oblivious;
There is no part information and thus topology (e.g., connection) and structure (e.g, symmetry)
are not well-defined. To learn a structure-aware model, we need to look into
neural models for structural representations such as graphs.

\paragraph{Generative neural models for structures.}
To learn feature representation for general structures, other than 1D or 2D grids,
some works attempted to extend convolutional neural networks to graph data~\cite{henaff2015,duvenaud2015,niepert2016}.
However, it seems hard to utilize these networks to construct generative models since
it is unclear whether and how the feature encodings can be made invertible to support
structure generation.
Socher et al.~\shortcite{socher2011,socher2012} pursue a different approach to this problem,
by utilizing recursive neural networks (RvNN) which sequentially collapses edges of a graph,
yielding recursive structure encoding with binary trees.
Inspired by this work, Li et al.~\shortcite{li2017} learn a generative recursive auto-encoder
for 3D shape structures.
Our work adapts this model for indoor scene generation with non-trivial extensions.

\section{Overview}
\label{sec:overview}



Our RvNN-VAE framework for generating 3D indoor scenes is trained on a large 3D scene dataset,
where each scene is composed of a set of labeled objects which are represented by bounding
boxes (OBBs). Once trained, the RvNN-VAE is used to generate new scenes through decoding a randomly sampled noise vector into a hierarchy of OBBs with object labels. The labeled OBBs are then replaced with 3D objects retrieved from a 3D shape database based on object category and model dimensions. Figure \ref{fig:network_struct} shows the high-level architecture of the RvNN-VAE and Figure \ref{fig:pipeline} illustrates the generative process enabled by the learned deep neural network.

\begin{figure}[t!]
	\includegraphics[width=\linewidth]{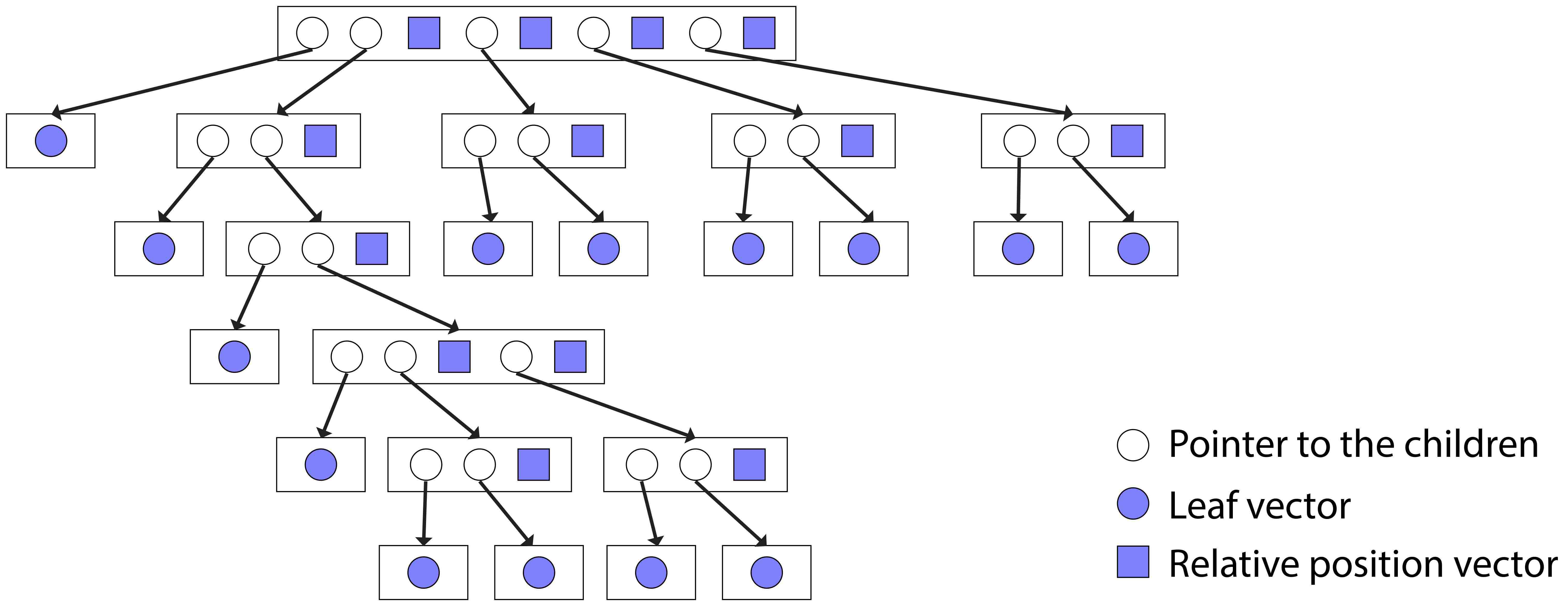}
	\caption{A hierarchical {\em vectorized\/} encoding for an indoor scene. Leaf vectors record object sizes and labels; internal nodes contain positional information of the child nodes {\em relative\/} to that of the first child node.}
	\label{fig:datastructure}
\end{figure}

\paragraph{Structural scene representation}
Given an input 3D indoor scene composed of a set of labeled objects, we organize the objects including walls and floor of a scene into a hierarchy, based on their spatial relations (Section \ref{sec:data}). Objects are present at the leaf level in the hierarchies, while each internal node represents a group of objects under its subtree. Figure~\ref{fig:datastructure} shows an illustration of hierarchical structure for indoor scenes. 
The object labels and sizes are encoded in the leaf vectors, and the spatial placement information is encoded by relative positions between sibling nodes in the hierarchy.

\paragraph{Recursive VAE} To learn the layout of 3D indoor scenes, we train a Variational Auto-Encoder (VAE) whose encoder maps an indoor scene or more specifically, its OBB hierarchy, into a fixed length root code in a bottom-up, recursive fashion and the decoder works inversely (Section \ref{sec:Netork description}). This is illustrated in Figure \ref{fig:network_struct}. 
During encoding, the box encoder is first applied to each of the leaf vectors, to map them into fixed length leaf codes. Then, the RvNN repeatedly invokes one of the four encoders and outputs a code until the code for the root node is generated. The decoder module performs an inverse process to reconstruct the hierarchy with the help of a node classifier that is trained simultaneously to predict the node type. 

\paragraph{Scene generation.} Once trained, the network learns a distribution of root codes that correspond to various scenes. Therefore, given a root code sampled from the distribution, the trained decoder module will decode the root code into a hierarchy of OBBs, as shown in Figure~\ref{fig:pipeline}. The decoded hierarchy includes the relative positions between sibling nodes and the bounding boxes in the leaf nodes.

\section{Structural Scene Representation}
\label{sec:data}

Indoor scenes are often characterized by their layouts and the classes of objects present in them, which we represent using labeled oriented bounding boxes (OBBs). For a generated scene to look realistic, it should follow some common 
object placement patterns within its subscenes. 
Object placement pattern involves both object classes and the relative positions between the associated objects. We make the key observation that such patterns are {\em hierarchical}, that is, they follow grouping rules at multiple levels of abstraction. Hence, we employ a hierarchical model to organize the objects in a scene and record their relative positions. 
Figure \ref{fig:datastructure} shows one such illustration. Here, when merging several nodes into internal nodes, we take into account their OBBs and the relative positions.

In this section, we describe the details of our structural scene representation, where our key innovation is the design of relative position encoding for object placements (Section~\ref{subsec:objpos}).


\subsection{Hierarchical scene structures}
Given an indoor scene, we organize the constituent objects into a hierarchy, where the objects and object groups are represented by the leaf nodes and internal nodes, respectively. We observe that the common features of indoor scenes lie within their corresponding subscenes, that include the objects and their placement w.r.t each other. 
To preserve these features, it is essential to build the hierarchies based on these sub-scenes. 

\paragraph{Pre-defining object relations}
Our idea is to first merge the commonly perceived ``relevant'' objects into sub-scenes and group the sub-scenes into a complete scene. However, because of the large variation in the training data in terms of number of objects, their geometry and placement across all the indoor scenes (even among scenes within the same room type), there's no consistent way of defining the sub-scenes. 
In spite of such a large variation, the relevant objects usually have consistent spatial relations among various scenes. This can be taken as a heuristic rule to construct the hierarchies for all the scenes. For example, a desk is likely to have a chair \textit{nearby}; a nightstand is likely to be placed \textit{adjacent to} a bed; the cabinets are placed \textit{against} the walls, etc.

Therefore, we build our hierarchies based on the spatial relations. To categorize these spatial relations, we characterize object-object relations into three categories: \textit{support}, \textit{surround} and \textit{co-occur}. The \textit{support} relation is defined for objects where one object is placed on top of the other. A set of objects is said to form a \textit{surround} relation if they have similar size and same label, and are located around a central object. The \textit{co-occurrence} relation is used as a ``catch-all'' relation. If two objects are not involved in a \textit{support} or a \textit{surround} relation, they are considered using the \textit{co-occurrence} relation. 
In our hierarchies, we treat the walls and the floor as special objects because they serve as the ``reference'' for the scene layout. In particular, we observe that the walls are responsible for the correct orientation of objects in a scene.

\begin{figure}[t]
\centering
\includegraphics[width=\linewidth]{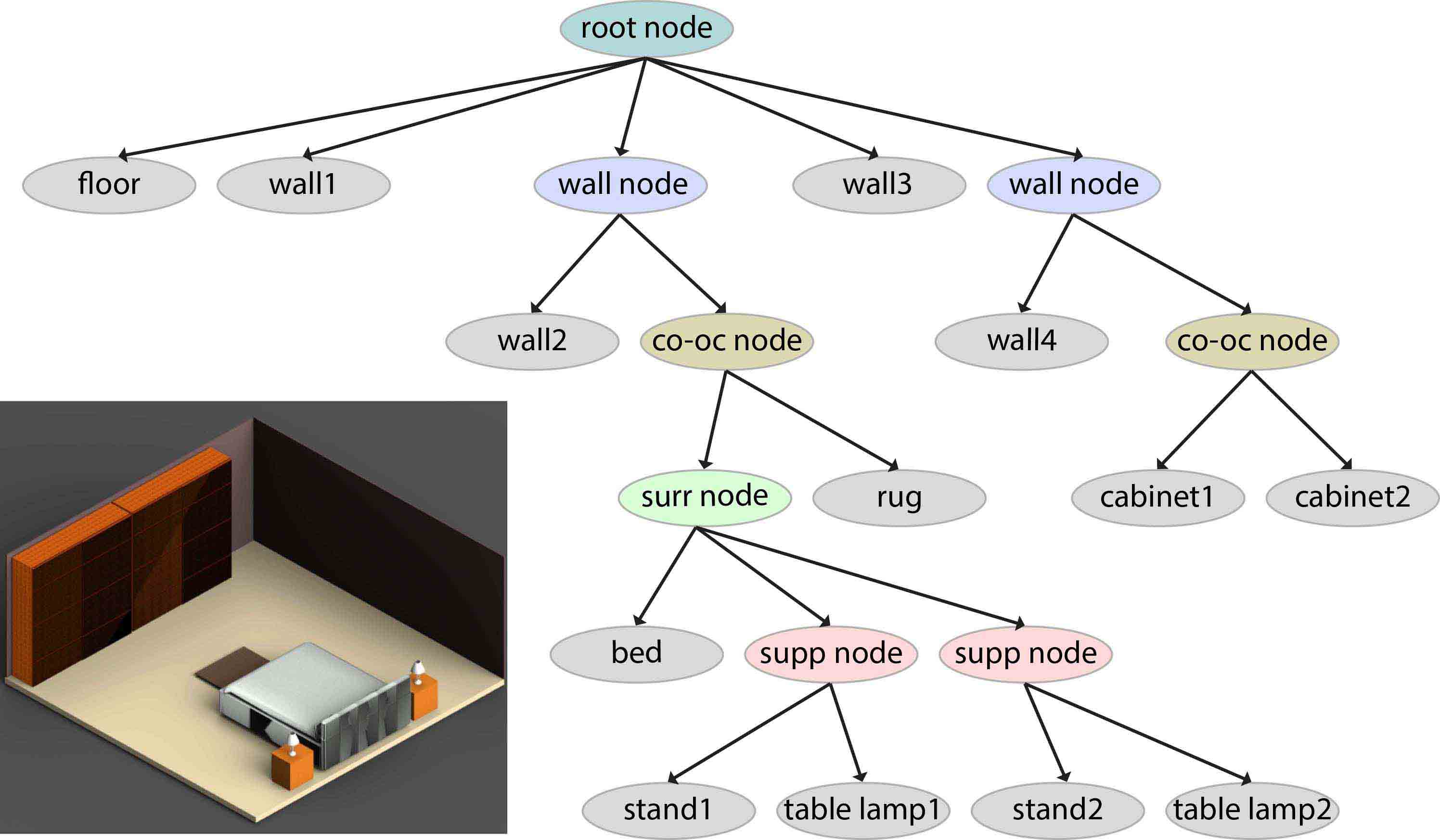}
\caption{A bedroom from the training set and the corresponding scene hierarchy. The root node always has a floor and four wall nodes as its children. Each wall is merged with its associated object clusters, which are organized into sub-trees based on finer-grained object-object relations.}
\label{fig:scene_hier}
\end{figure}

\paragraph{Building training scene hierarchies}
To build a scene hierarchy, we first cluster the objects in the scene based on the closest walls they are placed against. Then, we build a sub-tree using the object relations in every cluster. The \textit{support} relation has the highest priority, followed by the \textit{surround} relation and then the \textit{co-occurrence} relation. Figure~\ref{fig:scene_hier} shows a bedroom with various object relations and the corresponding scene hierarchy. Leaf nodes correspond to objects and non-leaf nodes represent object groups with certain relations. We observe that the two pairs of nightstands and table lamps are first merged by the \textit{support} relation independently, and then merged with the bed with the \textit{surround} relation.

To highlight these features to the network,  
we place the children of each kind of internal-node in a particular order. To elaborate, for a \textit{support} node, its second child (object or object group) is supported by the first child (object). 
 For a \textit{surround} node, its first child is the central object around which the other two objects (object groups) are placed. For a \textit{co-occurrence} node, the children are sorted based on the sizes of their OBBs -- the first child having the largest OBB. These orders enhance the common features in the dataset, making it easier for the network to learn.

Once the object clusters are available, we merge the walls and the associated clusters to form the ``wall nodes''. Then the floor and all the ``wall nodes'' are merged to get the ``root node''. Figure~\ref{fig:scene_hier} shows the overall hierarchy for a bedroom. In the hierarchy, we have different types of nodes (leaf nodes, support nodes, co-occurrence nodes, surround nodes, wall nodes and root node), corresponding to objects or object groups with different relations.

\subsection{Scene object representation}
\label{subsec:objpos}

We represent a 3D scene based on semantic and geometric information about the objects therein and
object-object relations, in a scene hierarchy, as shown in Figure~\ref{fig:datastructure}. Specifically, in
each leaf node, we record information about a 3D object, including the geometries (e.g., length, height,
etc.) of its 3D OBB, under a local frame, and the object's semantic label, encoded as a one-hot vector
whose dimension equals to the number of object categories for the corresponding scene type. In each internal node, we record the {\em relative\/} positioning between the OBBs of two sibling child nodes. It suffices to encode 2D layout information since object positioning along
the height dimension can be defined by the \textit{support} relation.
%
Overall, the relative (2D) object positioning is encoded in a 28-dimensional vector of real numbers and bits, as we explain below.

\begin{figure}[t!]
	\centering
	\includegraphics[width=0.5\linewidth]{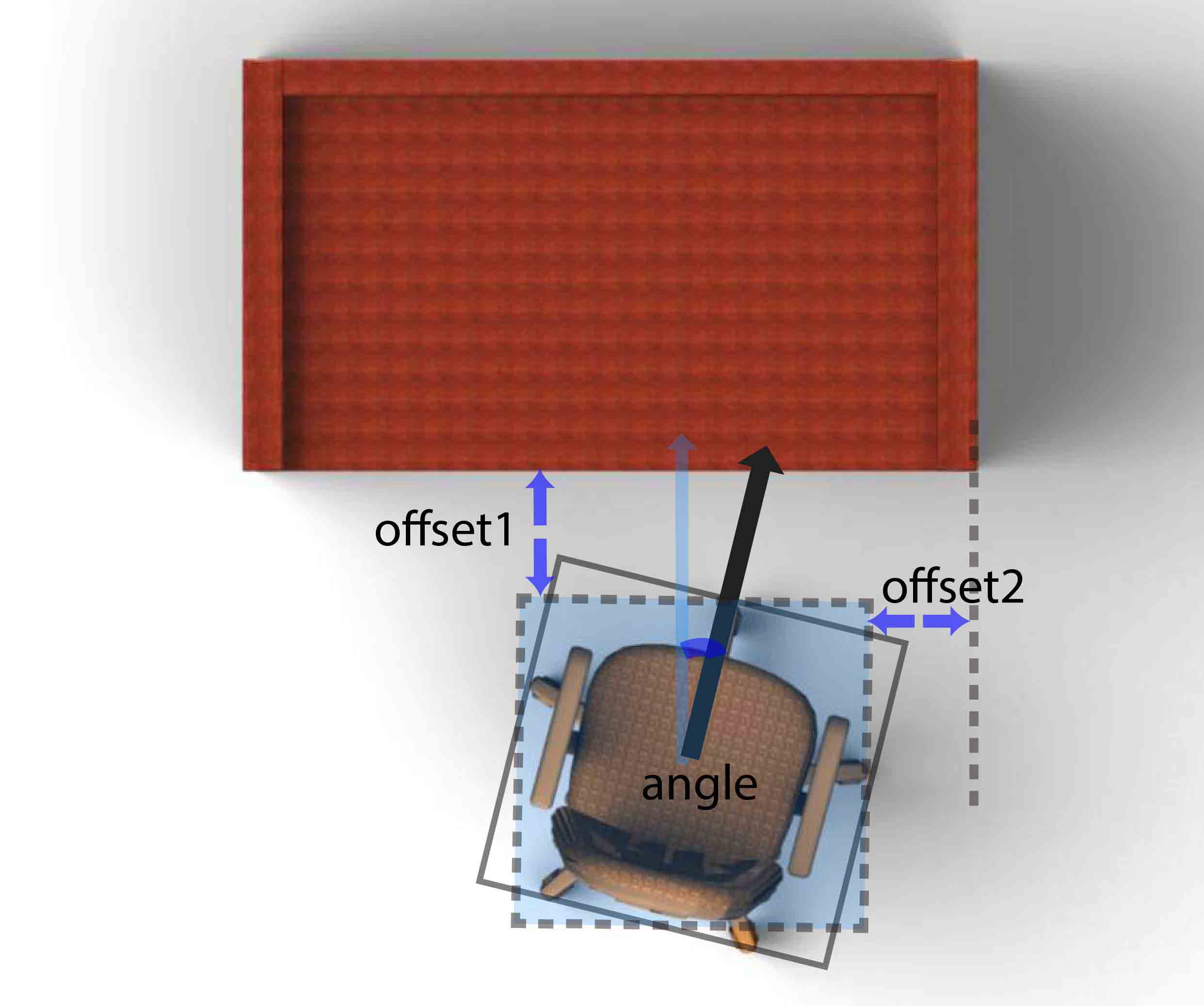}
	\caption{Defining relative positions between a desk and a chair using one rotation angle and two offsets between closest edges from the two OBBs.}
	\label{fig:relpos2d}
\end{figure}
\begin{figure}[t!]
	\centering
	\includegraphics[width=\linewidth]{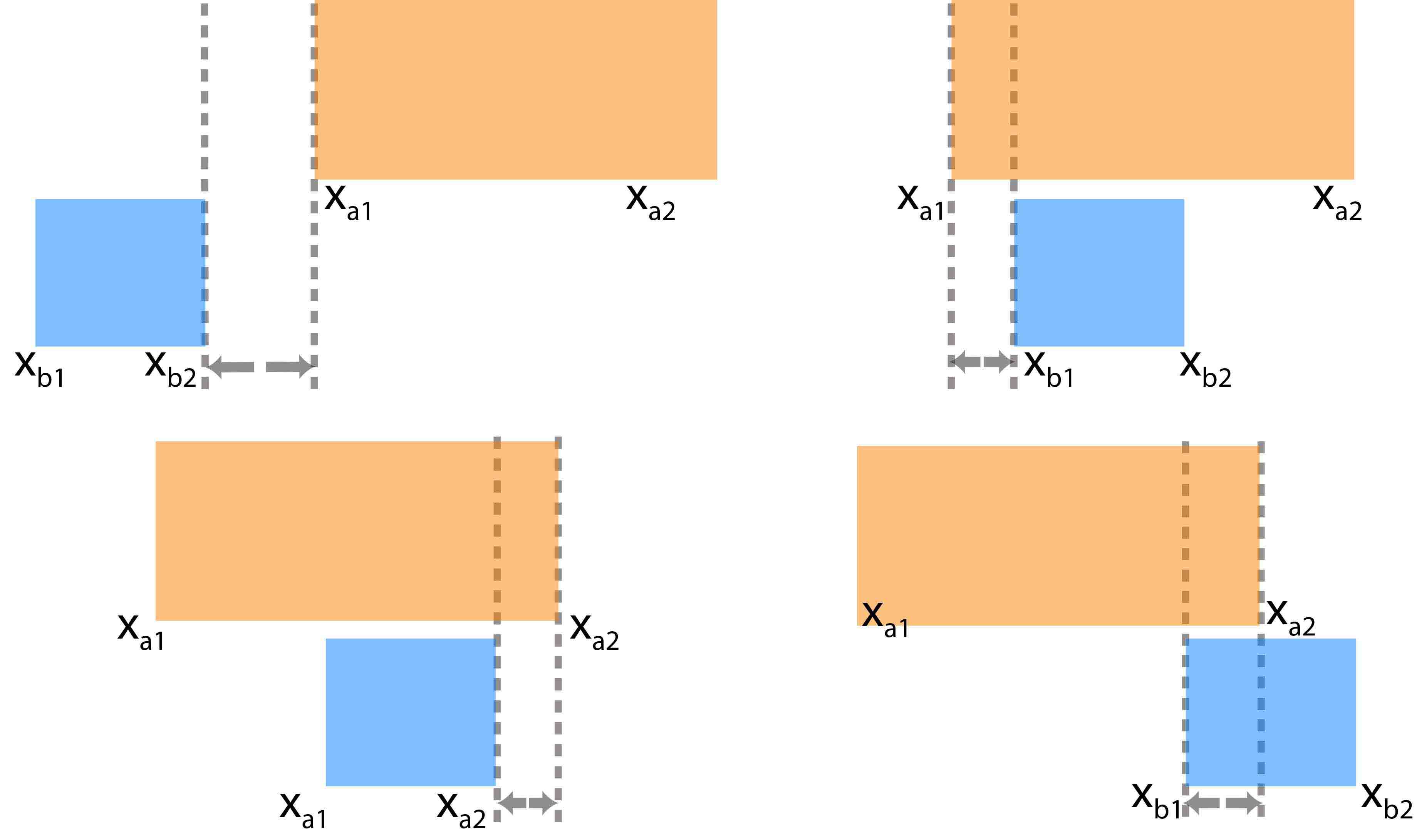}
	\caption{Along the horizontal direction, there are four cases of closest edge pairs from two axis-aligned boxes (blue and orange). The corresponding offset values are indicated by horizontal arrows.}
	\label{fig:relpos1d}
\end{figure}

\paragraph{Orientation and offsets}
To encode the relative positioning of one box, referred to as the target box, with respect to a reference box,
we first rotate the target box so that it is axis-aligned with the reference, where the {\em rotation angle\/} is stored
in the 28-D vector. Then we store {\em two offset values\/}, each representing the distance between two
{\em closest\/} edges from the two boxes, one along the horizontal direction and one along the vertical direction, as shown in Figure~\ref{fig:relpos2d}.
In Figure~\ref{fig:relpos1d}, we show that along the horizontal direction, there are 4 possible
cases for the pair of closest edges. Hence, we can use a $4\times 4=16$-bit ``two-hot'' indicator vector
to identify which two pairs of edges from the two boxes would define the two offset values.

\paragraph{Attachment and alignment}
Certain object-object attachment or alignment relations are rather strict, e.g., all beds have at least one side leaning against a wall. However, such strict relations are difficult to enforce precisely using only the angle and offset values since these values are easily affected by noise and one can always expect variations in them from the training data. To allow precise attachments and alignments in the generated scenes, we
opt to encode such relations explicitly using {\em bit codes\/} as a means of reinforcement beyond encoding object positions using offsets and angles.

Specifically, attachment is encoded by a $4$-bit one-hot vector indicating whether and how the two boxes are attached. The first bit is used to indicate whether or not the two boxes are attached, the next two bits indicate whether the boxes are attached along any one of the two axes, and the last bit indicates if the objects are aligned along both the axes. 
Note that an attachment implies that the two boxes are axis-aligned and that two of their edges either overlap or lie along a line, e.g., the left edge of the target box overlaps with the right edge of the reference box or the right edges of the two boxes lie along a line without overlapping.

Alignment is encoded by a $5$-bit one-hot vector to indicate whether the two boxes are oriented at an angle of $0^\circ$, $90^\circ$, $180^\circ$, or $270^\circ$ with respect to one another, or none of the above. In the latter case, note that the rotation angle is already stored in the 28-D vector. To summarize, the 28-D vector contains three real numbers (rotation angle and two offsets) and 25 binary indicators (16 bits for edge pair identification; 4 bits for attachment; 5 bits for alignment).

\section{Recursive Model of Indoor Scenes}
\label{sec:Netork description}

\begin{figure*}[h]
	\centering
	\includegraphics[width=\linewidth]{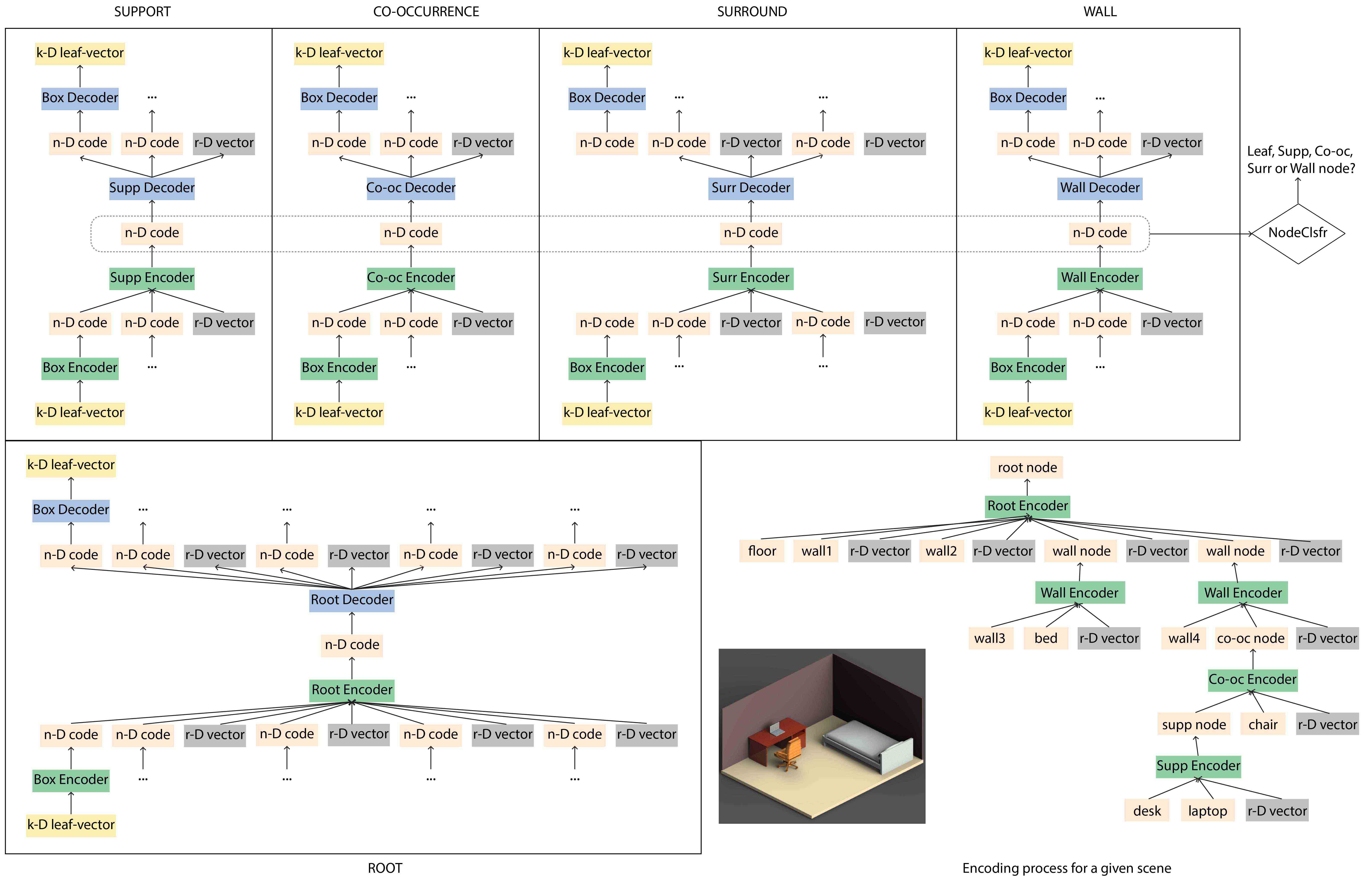}
	\caption{Autoencoder training setup. The r-D vectors are the relative position vector between the objects as explained in each Encoder/Decoder module. Ellipsis dots indicate that the code could be either the output of \textit{BoxEnc, SuppEnc, Co-ocEnc or SurrEnc}, or the inputs to \textit{BoxDec, SuppDec, Co-ocDec or SurrDec}. The encoding process for objects in a scene is shown on the bottom right.}
	\label{fig:training}
\end{figure*}

In this section, we describe our method to learn a representation of indoor scenes, of varying complexity, as fixed-dimensional codes. Our model is based on Recursive Autoencoders (RAE) for unlabeled binary trees, developed by Socher et al.~\cite{socher2011}. RAEs were originally intended for parsing natural language sentences in a discriminative setting, trained on unlabeled parse trees. It consists of an encoder neural network that takes two $n$-dimensional inputs and produces a single $n$-D output, and a decoder network that recovers two $n$-D vectors from a single $n$-D vector. Given a binary tree with $n$-D descriptors for leaves, the RAE is used to recursively compute descriptors for the internal nodes, ending with a root code. The root code can be inverted to recover the original tree using the decoder. With an additional classifier that helps reconstruct the original tree topology from a given root code, the decoding process does not require access to the input hierarchy. 

In our scenario,
the primary goal of the scene encoder is to encode the hierarchical organization of objects in a manner consistently reflecting common patterns found in various scenes.
We extend the original RAE framework for the task of scene generation to accommodate non-binary nodes (the \textit{surround} and \textit{root} nodes) and multiple encoder-decoder pairs.
In addition to the non-binary nodes, our hierarchies are built based on various object-object relations with different features, and therefore it is natural to use specific encoder and decoder pair for each relation.

\paragraph{Autoencoder model:} Our recursive autoencoder comprises six distinct encoder/decoder pairs. A \textit{box} encoder is used to convert the leaf vectors into codes. To handle the object-object relations, we have \textit{support}, \textit{co-occurrence} and \textit{surround} encoders. To group a cluster with its corresponding wall, we devise a \textit{wall} encoder. Finally, a \textit{root} encoder merges the four wall nodes with the floor. Figure \ref{fig:training} show the inputs and outputs of all these encoder/decoder pairs.

Below, we describe the encoders and decoders in detail. Each encoder and decoder (except those for input/output boxes) is a multi-layer perceptron (MLP)~\cite{Minsky1969,Werbos1974}, defined as a neural network with a finite series of fully-connected layers. Each layer $l$ processes the output $x_{l - 1}$ of the previous layer (or the input) using a weight matrix $W^{(l)}$ and bias vector $b^{(l)}$, to produce the output $x_l$. Specifically, the function is:
$$x_l \ = \ \tanh \left( W^{(l)} \cdot x_{l - 1} + b^{(l)} \right).$$
Below, we denote an MLP with weights $W = \{W^{(1)}, W^{(2)}, \dots\}$ and biases $b = \{b^{(1)}, b^{(2)}, \dots\}$ (aggregated over all layers), operating on input $x$, as $f_{W, b}(x)$. Each MLP in our model has one hidden layer.

\paragraph{{\sc {\bf Box}}}
The input to the recursive merging process is a collection of object bounding boxes plus the labels which are encoded as one-hot vectors. They need to be mapped to $n$-D vectors before they can be processed by different encoders. To this end, we employ an additional single-layer neural network \textit{BoxEnc}, which maps the $k$-D leaf vectors of an object (concatenating object dimensions and the one hot vectors for the labels) to a $n$-D code, and \textit{BoxDec}, which recovers the $k$-D leaf vectors from the $n$-D code. These networks are non-recursive, used simply to translate the input to the code representation at the beginning, and back again at the end. Each network's parameters comprise a single weight matrix, and a single bias vector.

\paragraph{{\sc {\bf Support}}}
The encoder for the support relation, \textit{SuppEnc}, is an MLP which merges codes of two objects (or object groups) $x_1$, $x_2$ and the relative position between their OBBs $r_{{x_1}{x_2}}$ into one single code $y$. In our hierarchy, we stipulate that the first child $x_1$ supports the second child $x_2$. The encoder is formulated as:
$$\textit{y} = f_{W_{Se}, b_{Se}} ([x_1 \  x_2 \  r_{{x_1}{x_2}}])$$
The corresponding decoder \textit{SuppDec} splits the parent code \textit{y} back to its children $x'_1$ and $x'_2$ and the relative position between them $r'_{{x_1'}{x_2'}}$, using a reverse mapping as shown below:
$$[x_1' \ x_2' \ r_{{x_1'}{x_2'}}'] = f_{W_{Sd}, b_{Sd}} \left( y \right)$$

\paragraph{{\sc {\bf Surround}}}
\textit{SurrEnc}, the encoder for the surround relation is an MLP which merges codes for three objects ${x_1},{x_2},{x_3}$ and two relative positions $r_{{x_1}{x_2}},r_{{x_1}{x_3}}$ into one single code $y$. In surround relation, we have one central object around which the other two objects are placed on either side. So we calculate the relative position for the two surrounding objects w.r.t the central object. This module can be extended to take more than three children nodes but in our case, we only consider three. The encoder is formulated as:
$$\textit{y} = f_{W_{SRe}, b_{SRe}} \left( [x_1 \  x_2 \  r_{{x_1}{x_2}} \ x_3 \   r_{{x_1}{x_3}}] \right)$$
The corresponding decoder \textit{SurrDec} splits the parent code \textit{y} back to children codes $x'_1$, $x'_2$ and $x'_3$ and their relative positions $r'_{{x_1'}{x_2'}},r'_{{x_1'}{x_3'}}$ using a reverse mapping, as shown below:
$$[x_1' \ x_2' \  r_{{x_1'}{x_2'}}'  \  x_3'  \  r_{{x_1'}{x_3'}}' ] = f_{W_{SRd}, b_{SRd}} \left( y \right)$$

\paragraph{{\sc {\bf Co-occurrence}}}
The encoder for the co-occurrence relation, \textit{Co-ocEnc}, is an MLP which merges codes for two objects (or object groups) $x_1,x_2$ and the relative position $r_{{x_1}{x_2}}$ between them into one single code $y$. In our structural scene hierarchies, the first child $x_1$ corresponds to the  object (or object group) with the larger OBB, than that of $x_2$. The \textit{Co-ocEnc} is formulated as:
$$\textit{y} = f_{W_{COe}, b_{COe}} \left( [x_1 \  x_2 \  r_{{x_1}{x_2}}] \right)$$
The corresponding decoder \textit{Co-ocDec} splits the parent code $y$ back to its children $x'_1$ and $x'_2$ and the relative position between them $r'_{{x_1'}{x_2'}}$, using a reverse mapping, given as:
$$[x_1' \ x_2' \ r_{{x_1'}{x_2'}}'] = f_{W_{COd}, b_{COd}} \left( y \right)$$

\paragraph{{\sc {\bf Wall}}}
The wall encoder, \textit{WallEnc}, is an MLP that merges two codes, $x_2$ for an object (or object group) and $x_1$ for the object/group's nearest wall, along with the relative position $r_{{x_1}{x_2}}$,
into one single code. In our hierarchy, a wall code is always the left child for the wall encoder, which
is formulated as:
$$\textit{y} = f_{W_{We}, b_{We}} \left( [x_1 \  x_2 \  r_{{x_1}{x_2}}] \right)$$
The corresponding decoder \textit{WallDec} splits the parent code \textit{y} back to its children $x'_1$ and $x'_2$ and the relative position between them $r'_{{x_1'}{x_2'}}$, using a reverse mapping, as shown below:
$$[x_1' \ x_2' \ r_{{x_1'}{x_2'}}'] = f_{W_{Wd}, b_{Wd}} \left( y \right)$$

\paragraph{{\sc {\bf Root}}}
The final encoder is the root module which outputs the root code. The root encoder, \textit{RootEnc}, is an MLP which merges codes for 5 objects $x_1,x_2,x_3,x_4,x_5$ and 4 relative positions $r_{{x_1}{x_2}},r_{{x_1}{x_3}}$, $r_{{x_1}{x_4}}$ and $r_{{x_1}{x_5}}$ into one single code. In our scene hierarchies, floor is always the first child $x_1$ and the remaining four child nodes correspond to four walls nodes in an anti-clockwise ordering (as seen from the top-view). The root encoder is formulated as:
$$\textit{y} = f_{W_{Re}, b_{Re}} \left( [x_1 \  x_2 \  r_{{x_1}{x_2}} \ x_3 \  r_{{x_1}{x_3}} \ x_4 \  r_{{x_1}{x_4}} \ x_5 \  r_{{x_1}{x_5}}] \right)$$
The corresponding decoder \textit{RootDec} splits the parent code \textit{y} back to child codes $x'_1$, $x'_2$ $x'_3$, $x'_4$ and $x'_5$ and the 4 relative positions $r'_{{x_1'}{x_2'}},r'_{{x_1'}{x_3'}}$, $r'_{{x_1'}{x_4'}}$, $r'_{{x_1'}{x_5'}}$  using a reverse mapping, as shown below:
$$[x_1' \ x_2' \  r_{{x_1'}{x_2'}}'  \  x_3'  \  r_{{x_1'}{x_3'}}' \ x_4'  \ r_{{x_1'}{x_4'}}' \ x_5' \ r_{{x_1'}{x_5}}'] = f_{W_{Rd}, b_{Rd}} \left( y \right)$$


The dimensions of the hidden and output layers for \textit{RootEnc} and \textit{RootDec} are 1,050 and 350, respectively. For the other modules, the dimensions are 750 and 250, respectively. This is because the root node has to accommodate more information compared to other nodes.

Lastly, we jointly train an auxiliary node classifier, \textit{NodeClsfr}, to determine which module to apply at each recursive decoding step. This classifier is a neural network with one hidden layer that takes as input the code of a node in the hierarchy, and outputs whether the node represents a box, support, co-occurrence, surround or wall node. Depending on the output of the classifier, either \textit{WallDec}, \textit{SuppDec}, \textit{SurrDec}, \textit{Co-ocDec} or \textit{BoxDec} is invoked.

\paragraph{Training}
To train our RvNN-VAE network, we randomly initialize the weights sampled from a Gaussian distribution. Given a training hierarchy, we first encode each leaf-level object using \textit{BoxEnc}. Next, we recursively apply the corresponding encoder (\textit{SuppEnc}, \textit{SurrEnc}, \textit{Co-ocEnc}, \textit{WallEnc} or \textit{RootEnc}) at each internal node until we obtain the root. The root codes are approximated to a Gaussian distribution by the VAE. A code is randomly sampled from this distribution and fed to the decoder. Finally, we invert the process, by first applying the \textit{RootDec} to the sampled root vector, and then recursively applying the decoders \textit{WallDec}, \textit{SuppDec}, \textit{SurrDec} and \textit{Co-ocDec}, followed by a final application of \textit{BoxDec} to recover the leaf vectors.

The reconstruction loss is formulated as the sum of squared differences between the input and output leaf vectors and the relative position vectors. The total loss is then formulated as the sum of reconstruction loss and the classifier loss, and Kullback-Leibler (KL) divergence loss for VAE.
We simultaneously train \textit{NodeClsfr}, with a five class softmax classification with cross entropy loss to 
infer the tree topology at the decoder end during testing.

\paragraph{Testing}
During testing, we must address the issue of decoding a given root code to recover the constituent bounding boxes, relative positions, and object labels. To do so, we need to infer a plausible decoding hierarchy for a new scene. To decode a root code, we recursively invoke the \textit{NodeClsfr} to decide which decoder to be used to expand the node. The corresponding decoder (\textit{WallDec}, \textit{SuppDec}, \textit{SurrDec}, \textit{Co-ocDec} or \textit{BoxDec}) is used to recover the codes for the children nodes until the full hierarchy has been expanded to the leaves. Since the leaf vectors output by \textit{BoxDec} are continuous, we convert them to one-hot vectors by setting the maximum value to one and the rest to zeros. In this way, we decode the root code into a full hierarchy and generate a scene based on the inferred hierarchy. 

\paragraph{Constructing the final scene} After decoding, we need to transform the hierarchical representation to 3D indoor scenes. With the object information present in leaf nodes and relative position information at every internal node, we first recover the bounding boxes for the non-leaf (internal) nodes in a bottom-up manner. By setting a specific position and orientation for the floor, we can compute the placement of the walls and their associated object groups, aided by the relative position information available at every internal node. When we reach the leaf level with a top-down traversal, we obtain the placement of each object, as shown in Figure \ref{fig:pipeline}(right most). The labeled OBBs are then replaced with 3D objects retrieved from a 3D shape database based on object category and model dimensions.

\section{Results, evaluation, and applications}
\label{sec:exp}

In this section, we explain our experimental settings, show results of scene generation, evaluate our method over different options of network design and scene representation, and make comparisons to close alternatives. The evaluation has been carried out with both qualitative and quantitative analyses, as well as perceptual studies.
Finally, we demonstrate several applications.

\begin{table}
	\begin{tabular} {| c | r | c |}
		\hline
		\textbf{Room type}& \textbf{\# Scenes}& \textbf{\# Object categories}\\
		\hline
		Bedroom&18,763&20\\
		\hline
		Living room&4,440&24\\
		\hline
		Kitchen&5,974&14\\
		\hline
		Office&3,774&16\\
		\hline
	\end{tabular}
\vspace{3pt}
	\caption{Some statistics for the training dataset used in our experiments. The object categories included in the table cover the most frequently appearing objects in SUNCG. The remaining, infrequent, objects were removed.}
	\label{tab:data}
\vspace{-15pt}
\end{table}

\paragraph{Dataset and training.}
The training data for our method is extracted from SUNCG~\cite{song2016ssc}, a large and diverse
collection of realistic indoor scenes encompassing scene types such as bedroom, living room,
kitchen, and office. In our experiments, we only consider rooms with a floor and rectangular wall
boundaries, without considering wall-mounted objects. Also, to facilitate the network training, we
remove the most infrequently appearing object categories over all scenes in every room type, as well as rooms with
too few or too many objects, resulting in a training set with statistics shown in Table~\ref{tab:data}.
Note that such a filtering of the scene dataset for network training is fairly standard, e.g., see~\cite{wang2018}.
Our training is based on PyTorch \cite{paszke2017automatic}, using Adam optimizer \cite{kingma2014adam} and batch normalization. For each room
type, we adopt a batch size that is $1/10$ the size of the training set and we train for 500 epochs. The code is available on the project page.

\begin{figure}[!t]
	\centering
	\includegraphics[width=\linewidth]{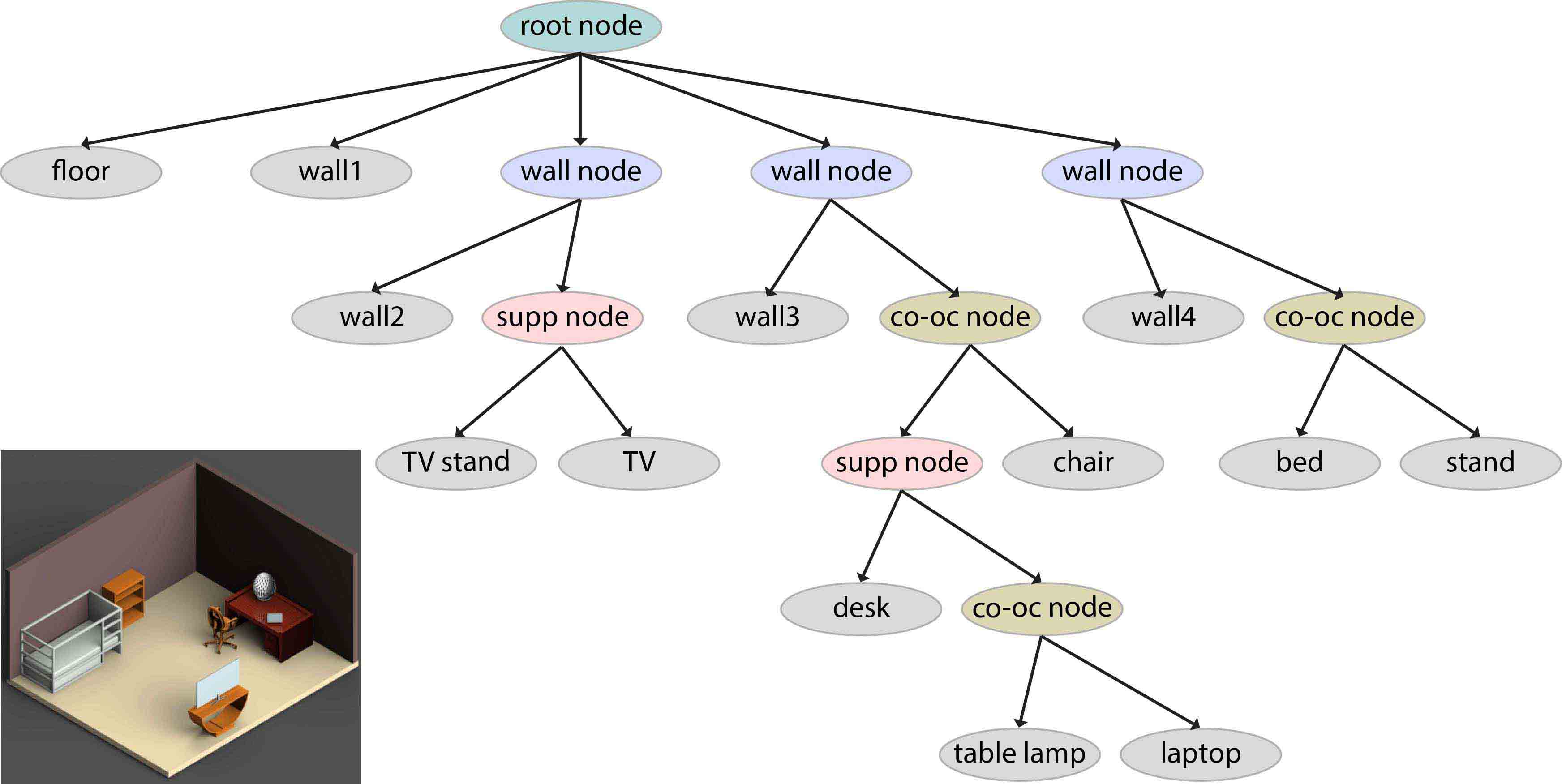}
	\caption{A scene generated by our method and its corresponding decoded hierarchy, which defines the scene structure.}
	\label{fig:decoded_hierarchy}
\end{figure}

\subsection{Scene generation results}
\label{subsec:scene_gen}

Using our method, a 3D indoor scene can be generated from a random vector by our trained RvNN-VAE network. The vector is first decoded into a structural scene hierarchy (e.g., see Figure~\ref{fig:decoded_hierarchy}),
and then converted into a scene via object retrieval. We shall demonstrate that the scenes generated by our method are plausible and diverse, while exhibiting some levels of novelty.

\begin{figure}[!t]
	\includegraphics[width=0.95\linewidth]{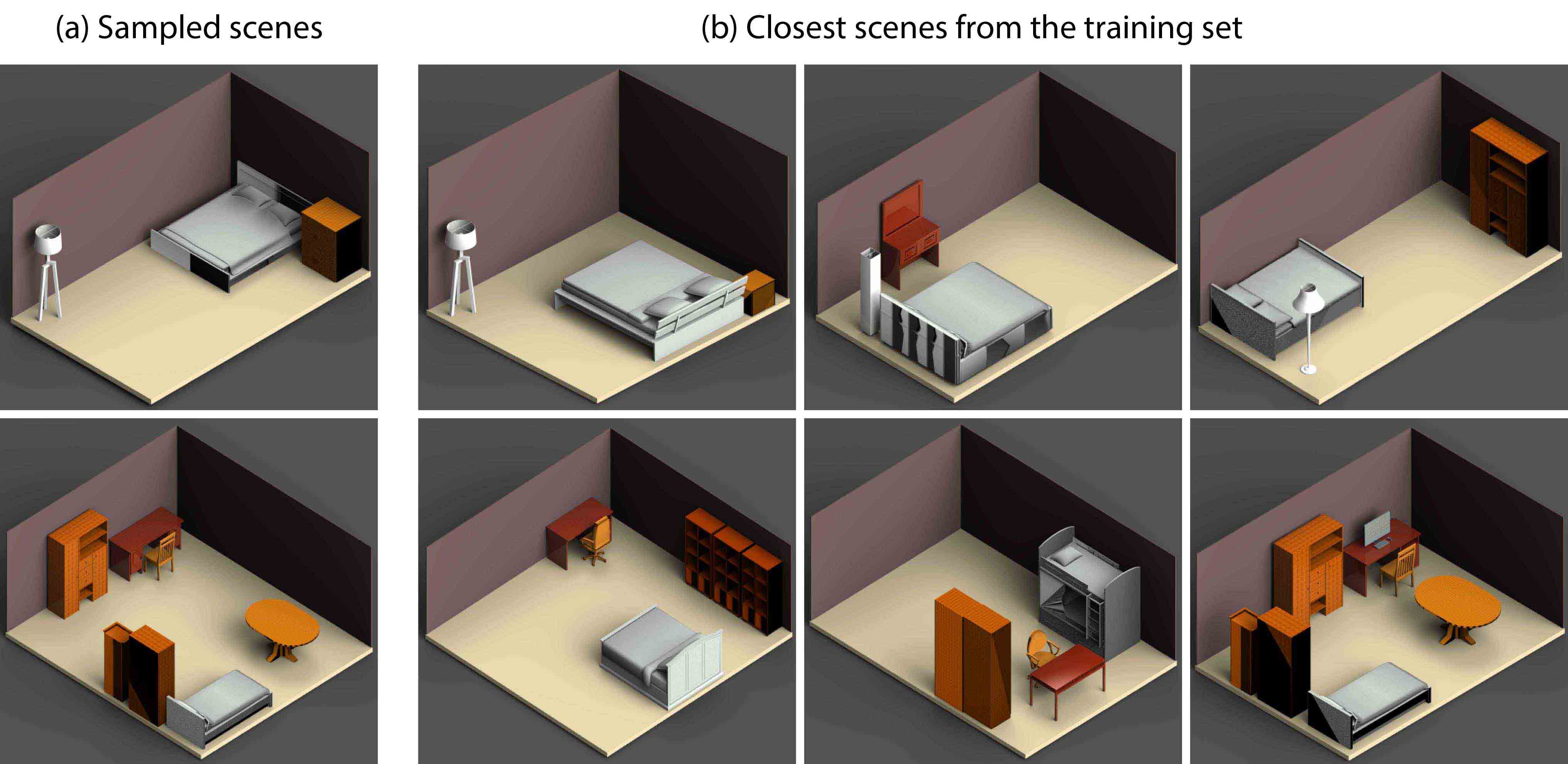}
	\caption{Top three closest scenes from the training set, based on graph kernel, showing diversity of scenes from the training set.} \label{knn_trainingset}
\end{figure}

To assess the diversity and novelty of the scenes, we need a scene similarity measure. To this end, we
adopt \textit{graph kernels}~\cite{fisher2011}, whose definition depends on object labels, sizes,
and object-object relations in a scene. 
Figure \ref{knn_trainingset} shows the three nearest neighbors, via graph kernels, in the training dataset to a given scene. These results are quite representative and indicative of the diversity of the training scenes, which
makes the learning problem more challenging.

\begin{figure*}[!t]
	\includegraphics[width=0.95\linewidth]{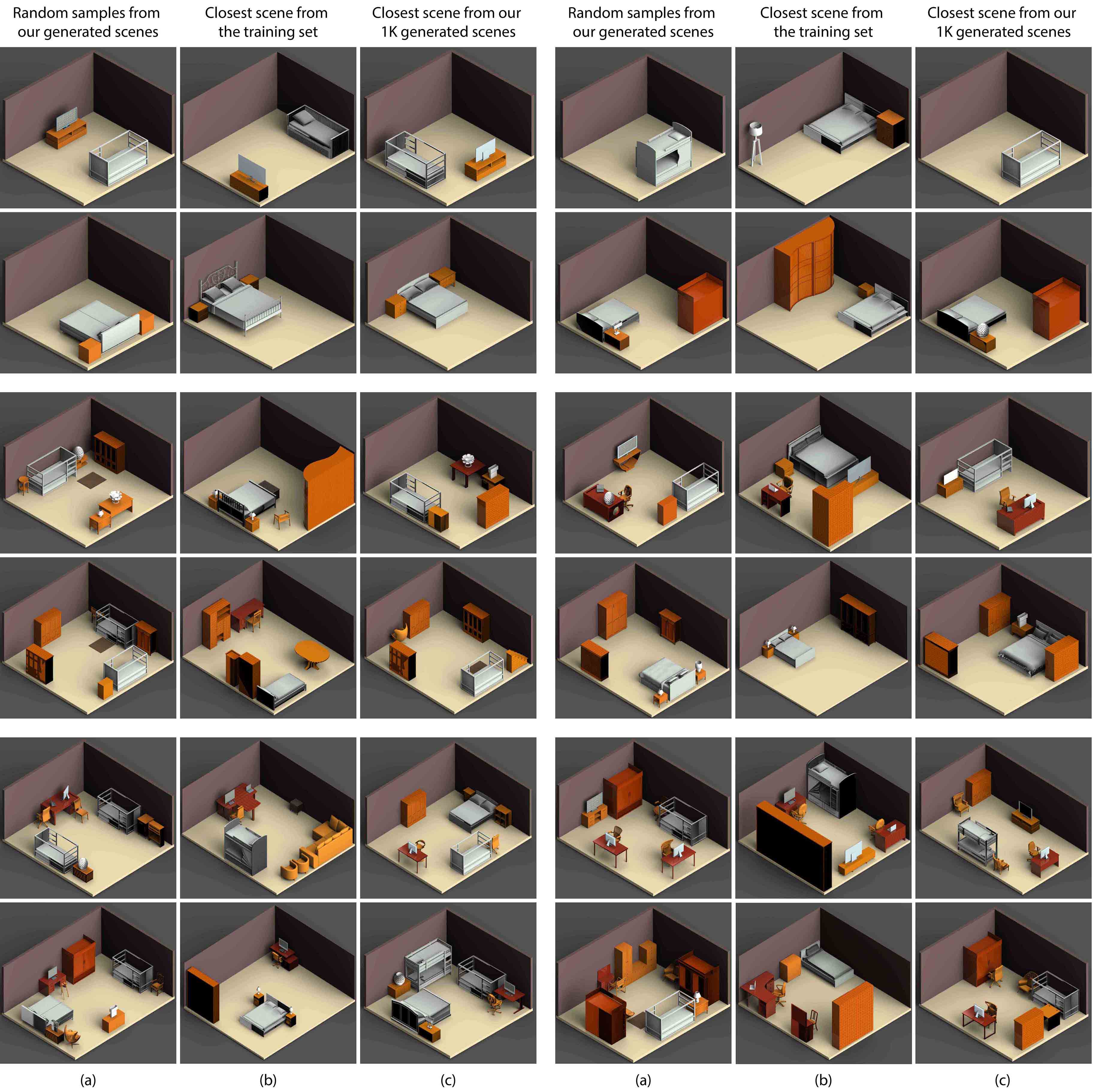}
	\caption{Bedrooms generated by our method (a), in comparison to (b) closest scene from the training set, to show novelty, and to (c) closest scene from among 1,000 generated results, to show diversity. Different rows show generated scenes at varying complexity, i.e., object counts.} \label{fig:gallery}
\end{figure*}

\begin{figure}[!t]
	\centering
	\includegraphics[width=0.95\linewidth]{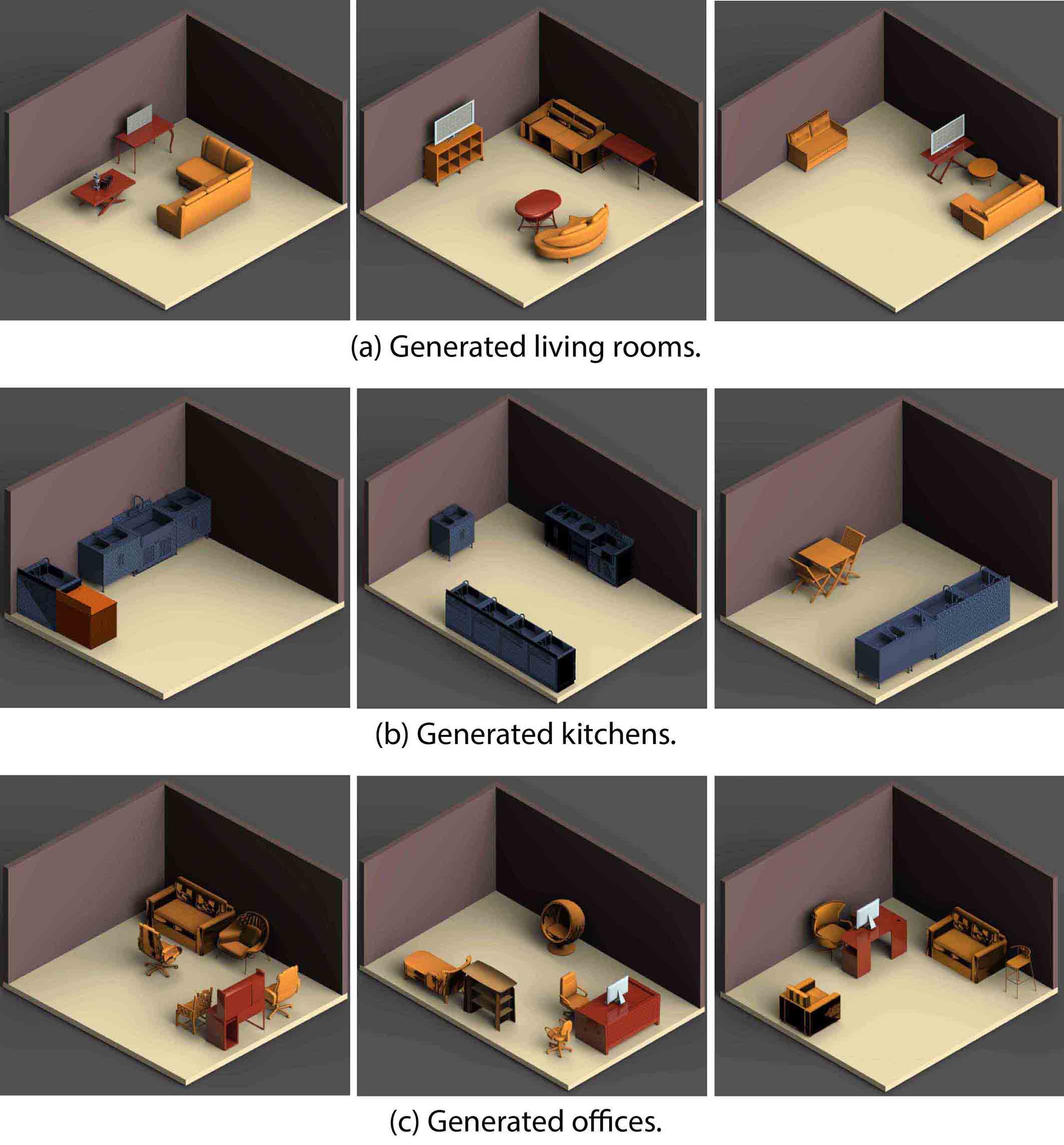}
	\caption{Several living room, kitchen, and office scenes generated by our method. More results can be found in the supplementary material.}
	\label{fig:morerooms}
\end{figure}

\paragraph{Scene generation}
Figure \ref{fig:gallery} presents a sampling of bedrooms generated by our method, which were randomly selected from a pool of 1,000 generated scenes, over three levels of scene complexity based simply on object counts. We can first observe that these scenes are plausible as exemplified by proper object placements against the walls and frequent object co-occurrences, e.g., TV+TV stand, bed+nightstands, desk+chair+computer combinations, etc. To demonstrate that our generative network does not simply memorize training scenes, we show the closest scene, based on graph kernels, from the training set. Also, we show the closest scene from the pool of generated scenes to assess diversity of the scenes generated.
Figure~\ref{fig:morerooms} shows several scenes generated for the other room types. More results can be found in the supplementary material.


\paragraph{Timing}
After training, our method can generate a 3D scene in a fraction of a second.
In comparison, recent work by Wang et al.~\shortcite{wang2018} reports a generation time of
4 minutes. More specifically, our generative network was able to produce 10K
bedroom scene hierarchies in a total of 94 seconds on a GPU. Then, it took 933 seconds to
convert the hierarchies to 3D scenes with object placements in MATLAB running on a CPU. Hence,
it takes {\bf 0.1027s} to generate a 3D bedroom scene, on average.
The timings are reported on a machine with one NVIDIA GeForce GTX 1080 Ti GPU and an Intel(R)
Core(TM) i7-8700 CPU @ 3.20GHz, with 64GB of memory. 
The training time depends on the size of the dataset and scene complexity. Running 500 epochs,
it took 36 hours to train the RvNN-VAE for bedrooms and about 7-12 hours for the other room
types.

\paragraph{Object co-occurrence}

\begin{figure}[!t]
	\centering
	\includegraphics[width=0.98\linewidth]{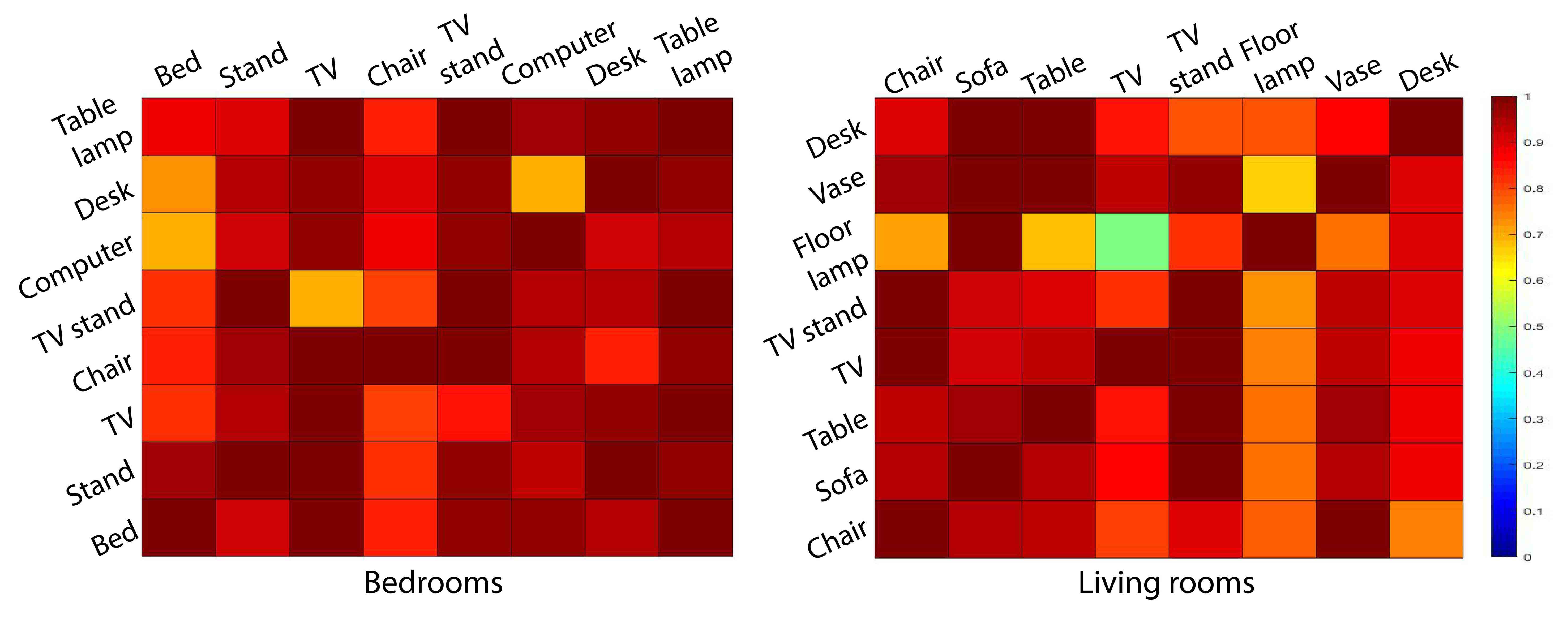}
	\caption{Similarity plots between object co-occurrences in the training set vs.~in the RvNN-generated
		scenes. Left: bedrooms; right: living rooms. Higher levels of similarity correspond to warmer colors. Note that we only plot similarities for pairs of categories with sufficiently high levels of co-occurrences; category pairs that rarely co-occur are discarded.}
	\label{fig:cooccur-matrix}
\end{figure}
One possible way to validate our learning framework is to examine its ability to reproduce the
probabilities of object co-occurrences in the training scenes. For this purpose, we define a conditional
probability for two object categories $c_1$ and $c_2$:
$P(c_1|c_2)=\frac{N(c_1,c_2)}{N(c_2)}$, where $N(c_1,c_2)$ is the number of scenes with
\textit{at least} one object belonging to category $c_1$ and one object belonging to category $c_2$,
and $N(c)$ is the number of scenes with \textit{at least} one object belonging to category $c$.
Figure~\ref{fig:cooccur-matrix} plots similarities between the $P(c_1|c_2)$'s over object category
pairs for bedrooms and living rooms, where the similarity is measured between the conditional
probabilities over the entire training set of scenes and $1,000$ randomly generated scenes of the
same room type. With the exception of the TV-floor lamp pair for living rooms, the strong
similarities indicate that our learned RvNN is able to generate scenes with similar object
co-occurrences compared to the training scenes.

\subsection{Comparisons}
\label{subsec:compare}

The most popular approach to 3D scene synthesis has so far been based on probabilistic graphical models~\cite{fisher2012, kermani2016learning, qi2018human}, 
which learn a joint conditional distribution of object co-occurrence from an exemplar dataset
and generate novel scenes by sampling from the learned model.

\begin{figure}[!t]
	\centering
	\includegraphics[width=0.99\linewidth]{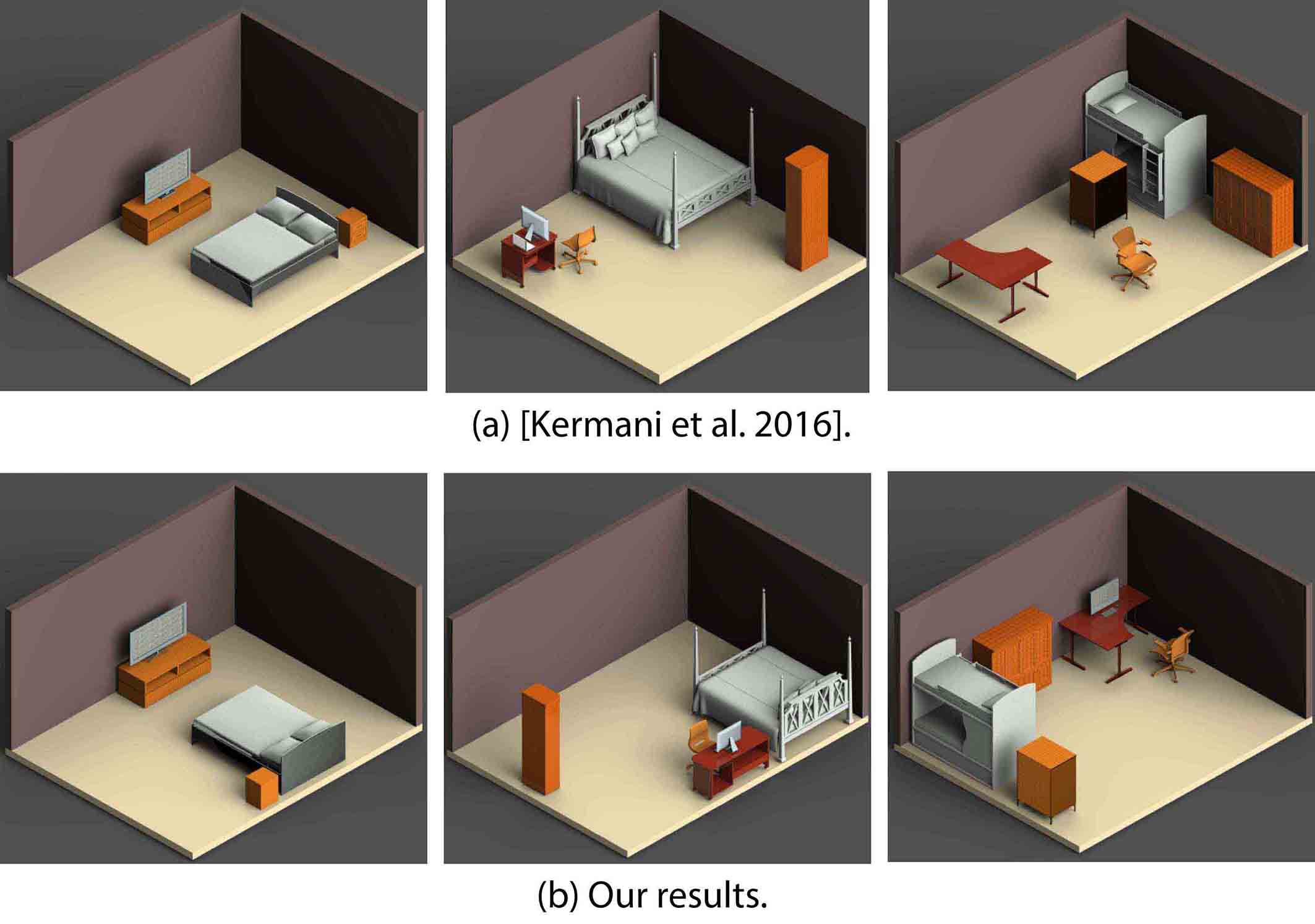}
	\caption{Comparison of scene generation with \cite{kermani2016learning}.}
	\label{comparison}
\end{figure}

To compare with the state-of-the-art, we take the closest work of Kermani et al.~\shortcite{kermani2016learning} which does not rely on human-activities~\cite{qi2018human} or
other auxiliary priors. In order to make a fair comparison, we implement their algorithm on the same set of scenes from SUNCG dataset that were used to train our RvNN-VAE network. In Figure \ref{comparison}, we present three randomly selected synthesized scenes using their approach and ours, with the same set of object models.  
Their method generates a valid set of objects but produces many
invalid arrangements and object orientations, affecting the plausibility of synthesized scenes. 
These are a result of their relative position encoding scheme and the underlying Monte-Carlo Markov Chain (MCMC) sampling model, which requires thousands of iterations to generate one plausible scene.
Another drawback of using sampling based techniques for scene synthesis is that there is no guarantee on \textit{exact alignment} between objects in the synthesized scenes, unlike ours.  
While our method generates large number of plausible, diverse scenes in seconds (e.g., $10$K bedroom scenes in $94$ seconds), it takes anywhere from a few minutes to tens of minutes to synthesize one scene (depending upon the complexity) using their approach.


To better compare the two methods, we perform a subjective analysis over the scenes generated by both methods, as well as those from the training set. In particular, we design two perceptual studies. 
In the first study, we ask the participants to select from a triplet of scenes the most plausible one. The triplet consists of scenes from the training set and results generated by the two methods (ours and \cite{kermani2016learning}), presented in a random order.
In the second study, the same set of participants are asked to rate each of the three scenes in the triplets based on their plausibility, on a scale of $1$ to $5$ ($5$ being most plausible).
The two studies are conducted separately, to minimize their mutual influence.  
The study employed $50$ participants with different backgrounds - $30$ graphics/vision researchers, $10$ non-graphics researchers and $10$ non-technical subjects. 
Each participant is presented with $10$ sets of triplets.
Therefore, we collect feedbacks from $50$ participants on $10 \times 3$ scenes for each study.

\begin{figure}[!t]
	\centering
	\includegraphics[width=\linewidth]{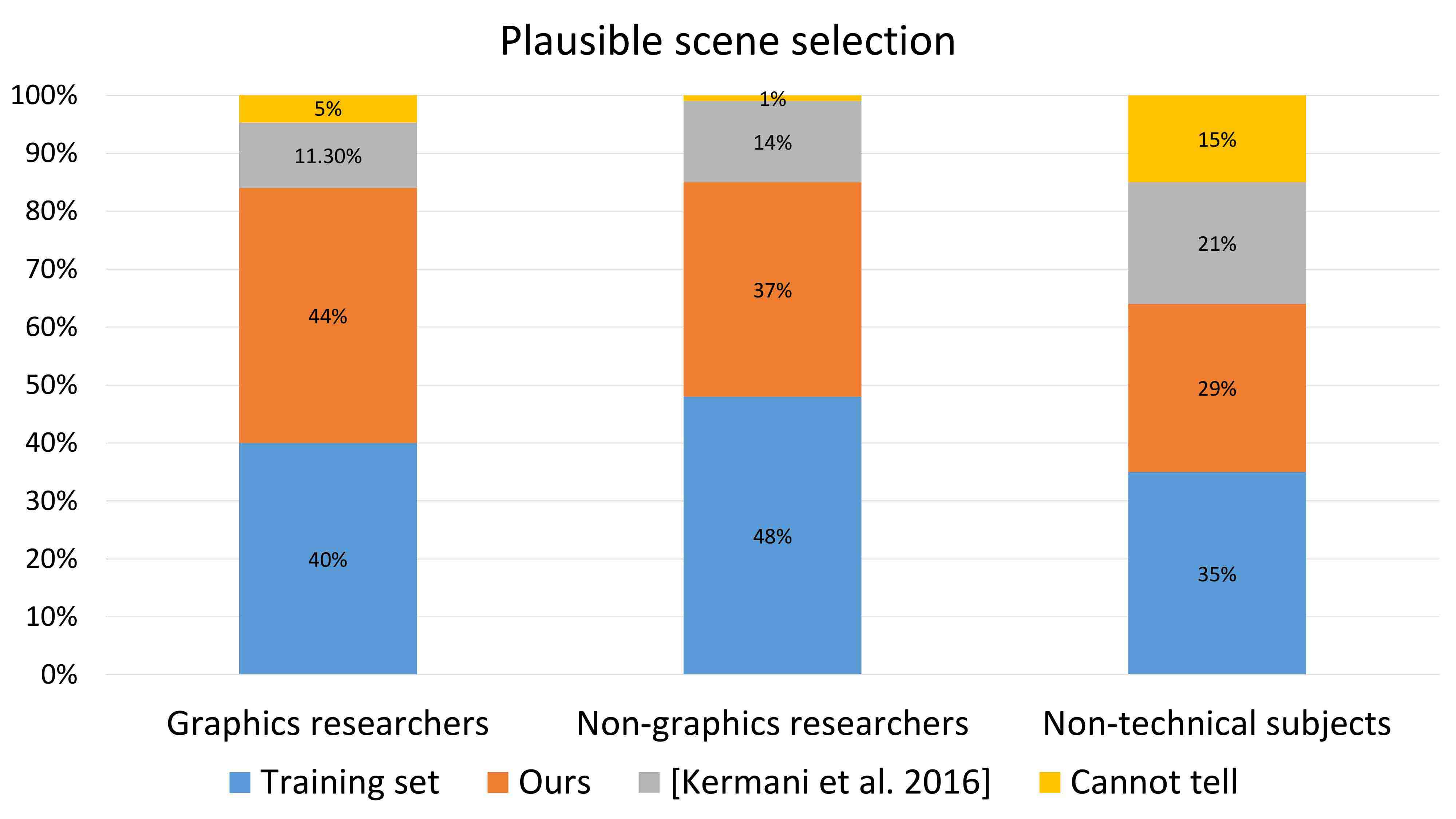}
	\caption{Statistics on plausible scene selection from the perceptual studies 
		: The participants are asked to select the most plausible scene for each triplet, consisting of bedroom scenes from the training set, our results and the results from \cite{kermani2016learning}. The plots show the percentage of 
		rated plausible scenes for each method.}
	\label{sceneselection}
\end{figure}

\begin{figure}[!t]
	\centering
	\includegraphics[width=\linewidth]{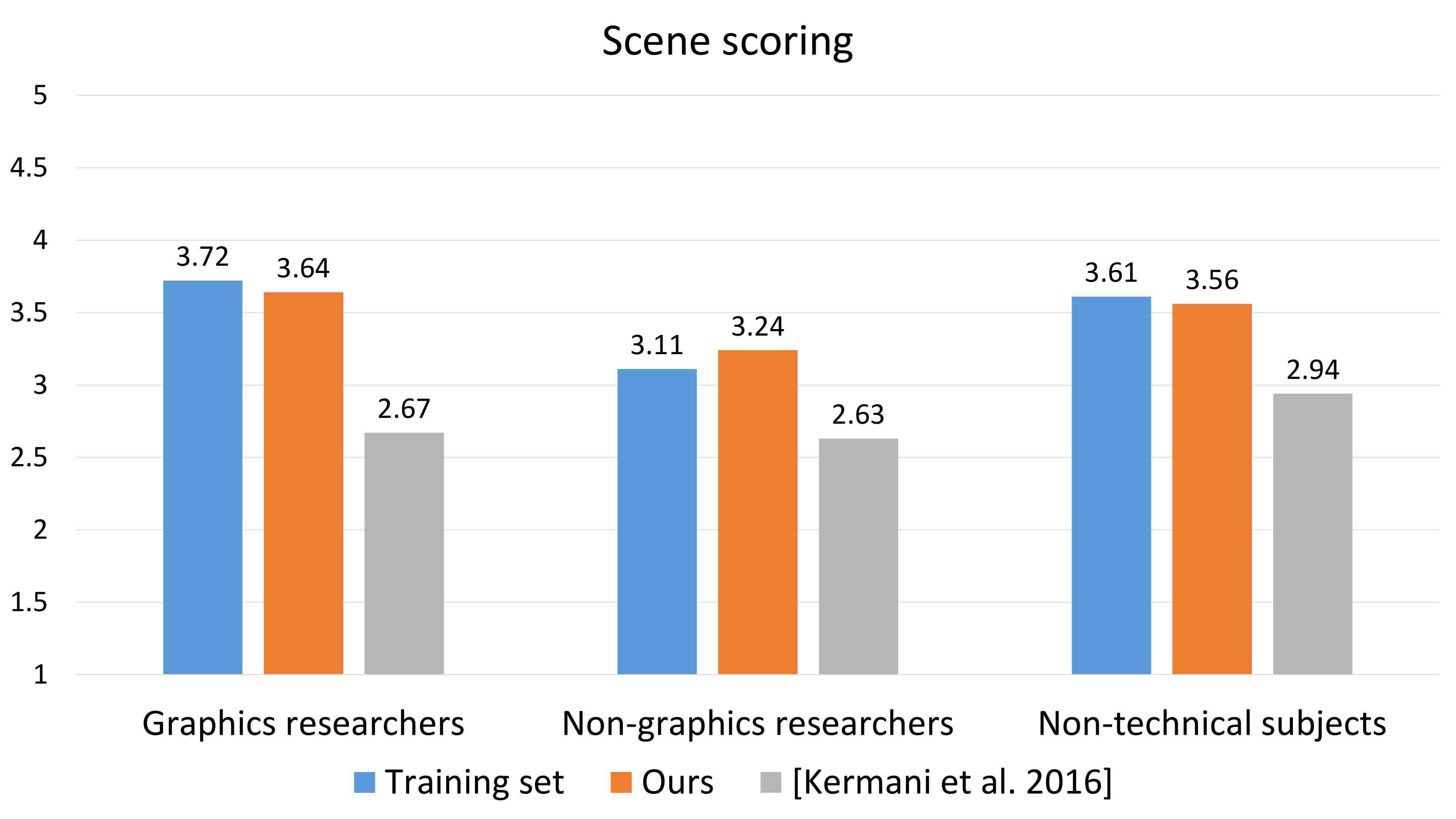}
	\caption{Results of plausibility rating by participants 
		for bedrooms: The participants are asked to rate a scene based on their plausibility on a scale of $1$ to $5$ (5 being the most plausible). The values are the average scores for scenes from the training set, our results and the results by \cite{kermani2016learning}.}
	\label{scenescoring}
\end{figure}

Figure \ref{sceneselection} and \ref{scenescoring} plot the statistics from the perceptual studies. 
From Figure \ref{sceneselection}, we see that the participants 
mostly think that the training scenes are more plausible,
since these scenes are human designed. The scenes generated by our method are also frequently chosen, especially by Graphics/Vision researchers, achieving a comparable level with training scenes.
Figure \ref{scenescoring} shows the statistics of plausibility rating.
We observe a comparable rating for scenes from the training set and our results,
demonstrating the high quality of our results.
We also ask for free-text responses from the participants to understand their basis of making a choice in the studies. From the feedback, participants 
without technical background tend to judge the scene plausibility based on their intuition on real world scenes. Therefore, they are more sensitive to object shapes (e.g., relative size) than object relations and overall layout. On the other hand, Graphics/Vision researchers pay more attention to object placements such as their orientation and their relative positions. This is why they tend to give slightly higher scores to all scenes. Moreover, non-(Graphics/Vision) researchers tend to give higher scores to our results even though they believed the scenes from the training set are generally more plausible than ours in the first study. 

\paragraph {Comparing scene plausibility to concurrent works.}
Concurrent works by Qi et al.~\shortcite{qi2018human} and Wang et al.~\shortcite{wang2018} are also trained on SUNCG for scene synthesis, but their scene generations are under certain constraints. Specifically, Qi et al.~\shortcite{qi2018human} use object sizes, object positions, orientations and human positions interacting with entities in a scene as inputs and rely on MCMC sampling to synthesize new scene layouts. On the other hand, the work by Wang et al.~\shortcite{wang2018} takes as input a partial scene geometry and synthesizes new layouts learned from CNNs trained on 2D top-view images of 3D scenes. Since GRAINS is an unconditional scene generator, we cannot directly compare it with these two works. Instead, we focus the comparison on {\em plausibility of the scene layouts\/} generated by the different methods, through pairwise perceptual studies.

\begin{figure}[!t]
	\centering
	\includegraphics[width=0.99\linewidth]{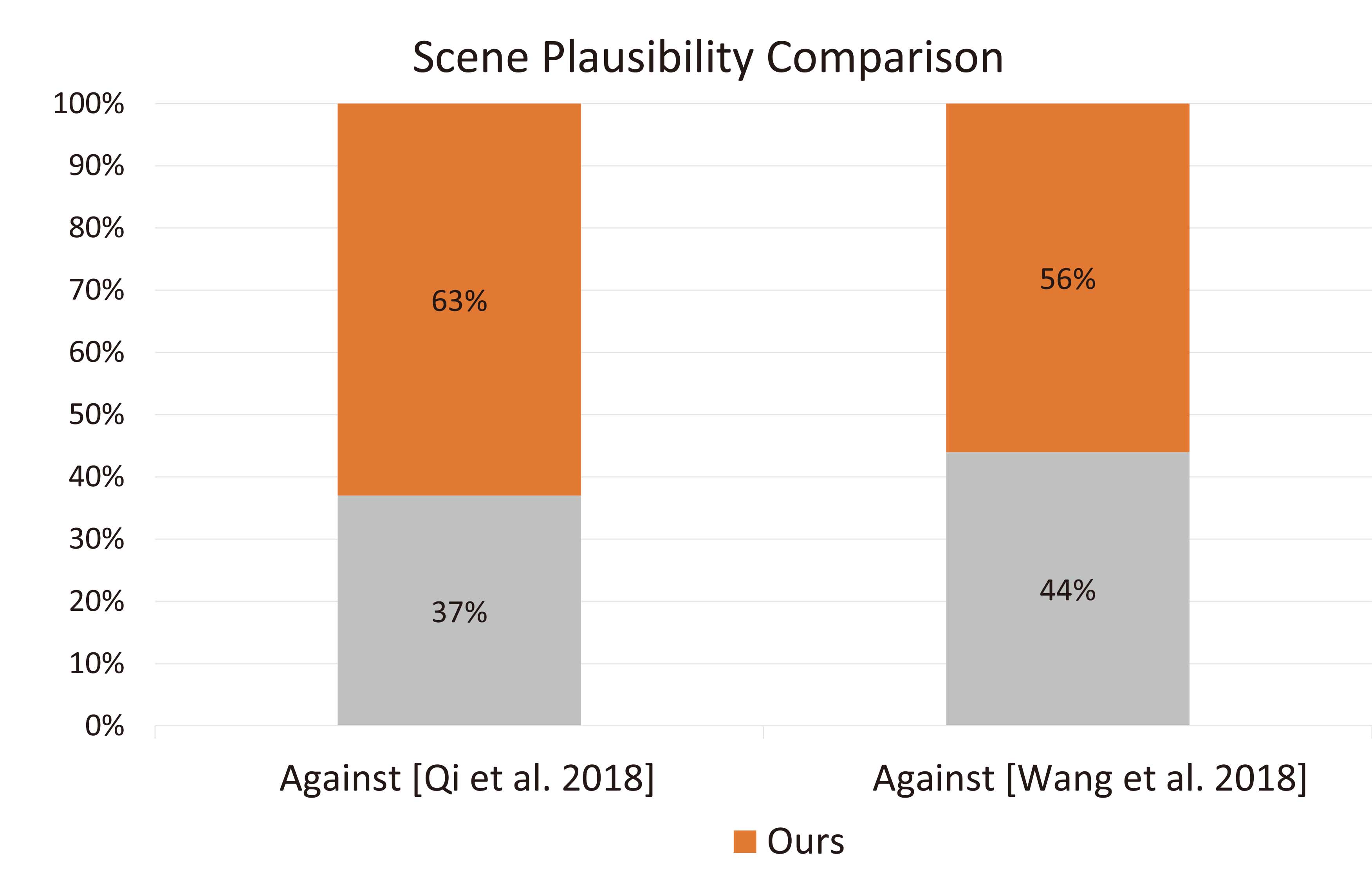}
	\caption{Percentage ratings in the pairwise perceptual studies, based on the votes for the plausibility comparisons between GRAINS and two state-of-the-art
			3D scene synthesis methods, namely~\protect\cite{qi2018human} and~\protect\cite{wang2018}.
			In each comparison, our method received a higher percentage of votes from 20 participants.}
		\vspace{-1.5em}
	\label{cvpr18_sig18_comparisons}
\end{figure}

To make the comparisons fair, we ensure that each pair of scenes compared contain the same or similar objects as follows. We take a scene generated by a competing method (either \cite{qi2018human} or \cite{wang2018}) and use it as a query to retrieve a scene generated (unconditionally) by GRAINS which most closely matches the query scene in terms of object counts over all object categories present in the query, e.g., number of tables, chairs, beds, etc. The query and retrieved scenes are paired up to set up a comparison. For consistency, we remove doors and windows from the generated scenes, if any. Furthermore, we adjust the 3D objects in the compared scenes so that they are either the same or similar in style. Hence, our focus is on comparing how well the methods were able to learn 3D scene layouts from (the same) SUNCG training data. Each comparative study consists of 15 pairs of scenes, with 5 pairs each from bedrooms, living rooms, and offices. Each pair of scenes is rendered from the same viewpoint and presented in a random order to 20 participants involved in the perceptual study.

In Figure~\ref{cvpr18_sig18_comparisons}, we show the percentages of participants based on how they voted for the plausibility comparisons between GRAINS and~\cite{qi2018human} and~\cite{wang2018}, respectively.
We can see that our method received 63\% of the total votes, over 37\% for Qi et al.~\shortcite{qi2018human}. This suggests that there is a higher uncertainty in the convergence of the scene layouts synthesized by their method, as it relies on MCMC sampling. It is also worth noting that it takes around ~22 minutes to synthesize one scene using their method as opposed to 102.7ms using GRAINS. 
%
The vote percentages for GRAINS and \cite{wang2018} are more comparable. However, aside from a gap in scene generation times (around 4 minutes by their method), we note that \cite{wang2018}
cannot produce scenes with 3D objects supported on top of other furniture, since their layout method was trained using top-view images. This may explain why GRAINS was slightly favored in the comparisons.

\subsection{Validation of network design and scene encoding}
\label{subsec:justify}

To validate several important choices in our network design and scene encoding, we show
experimental results with vs.~without these design decisions. Note that these choices also
highlight key differences between GRAINS and GRASS \cite{li2017}, which is the prior RvNN-VAE designed for
learning {\em shape\/} structures.

\begin{figure}[h]
	\centering
	\includegraphics[width=0.95\linewidth]{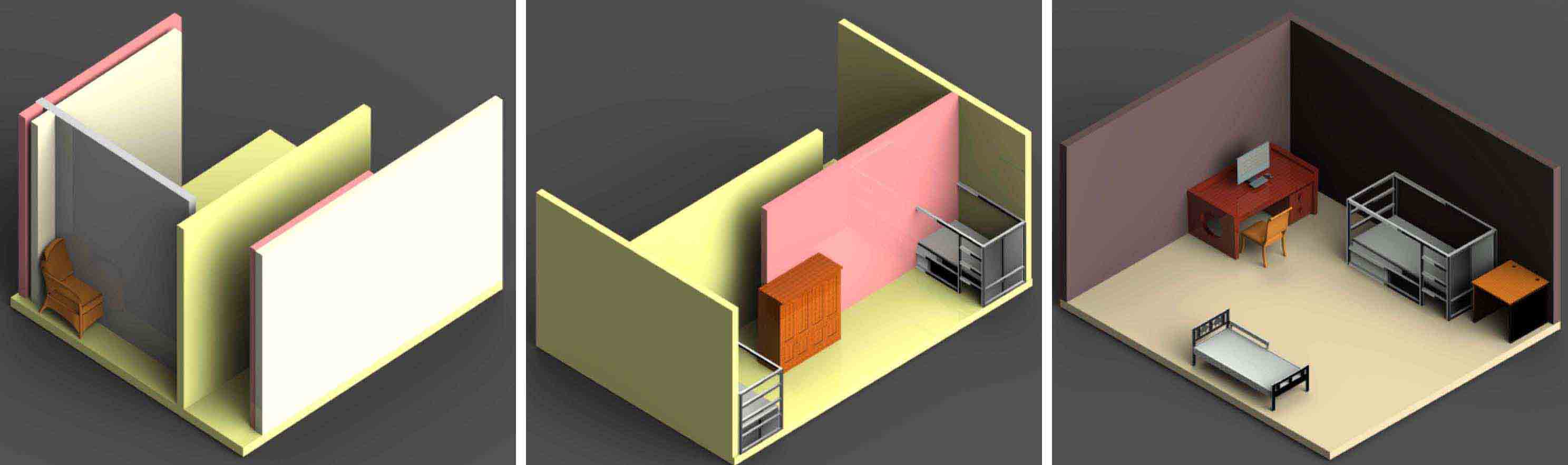}
	\caption{Typical bedroom scenes generated by our RvNN-VAE with various options involving wall
		and root encoders and decoders. (a) Without wall or root encoders and decoders. (b) With wall encoders
		and decoders, but not for root. (c) With both types of encoders and decoders.}
	\label{fig:wallroot_comp}
\end{figure}
\vspace{-1em}
\paragraph{Wall and root encoders/decoders}
By designating wall and root encoders and decoders with the last grouping operations in our RvNN and
carrying them out in a fixed order, we follow the premise that walls should serve a similar role as floors,
i.e., to ``ground'' the placements of objects in a room. Treating walls and floors just like any other
objects, i.e., without wall or root encoders and decoders, would lead to an RvNN that is much like
GRASS but with symmetry and assembly operations between shape parts replaced by object support,
co-occurrences, etc. Under this setting, walls are merely objects that have a high occurrence frequency.
It is difficult to train the network to always produce the correct number and
placements of walls, as shown by a typical result in Figure~\ref{fig:wallroot_comp}(a).
In fact, only 36.9\% of 1K generated scenes have exactly four walls, with most of these rooms having incorrect wall placements in the layout.


In Figure~\ref{fig:wallroot_comp}(b), we show a typical scene generated using our method where
wall encoders and decoders are incorporated, but not the root encoder. As we can observe, this time the
network is able to properly learn the positioning of objects against each wall. However, without
the root encoder, which accounts for and learns the relative positions of the four walls with respect to the
floor, the placements of the walls can be quite off. Without the designated root encoder, we must resort
to some alternative way to merge the walls and floors. When generating the result shown in Figure~\ref{fig:wallroot_comp}(b), we applied the wall encoders to merge two opposite walls, and then
the two pairs of walls. Finally, the combined node representing four walls is merged with the floor using a {\em support\/} relation.

\vspace{-0.5em}



\begin{figure}[!t]
	\centering
	\includegraphics[width=0.95\linewidth]{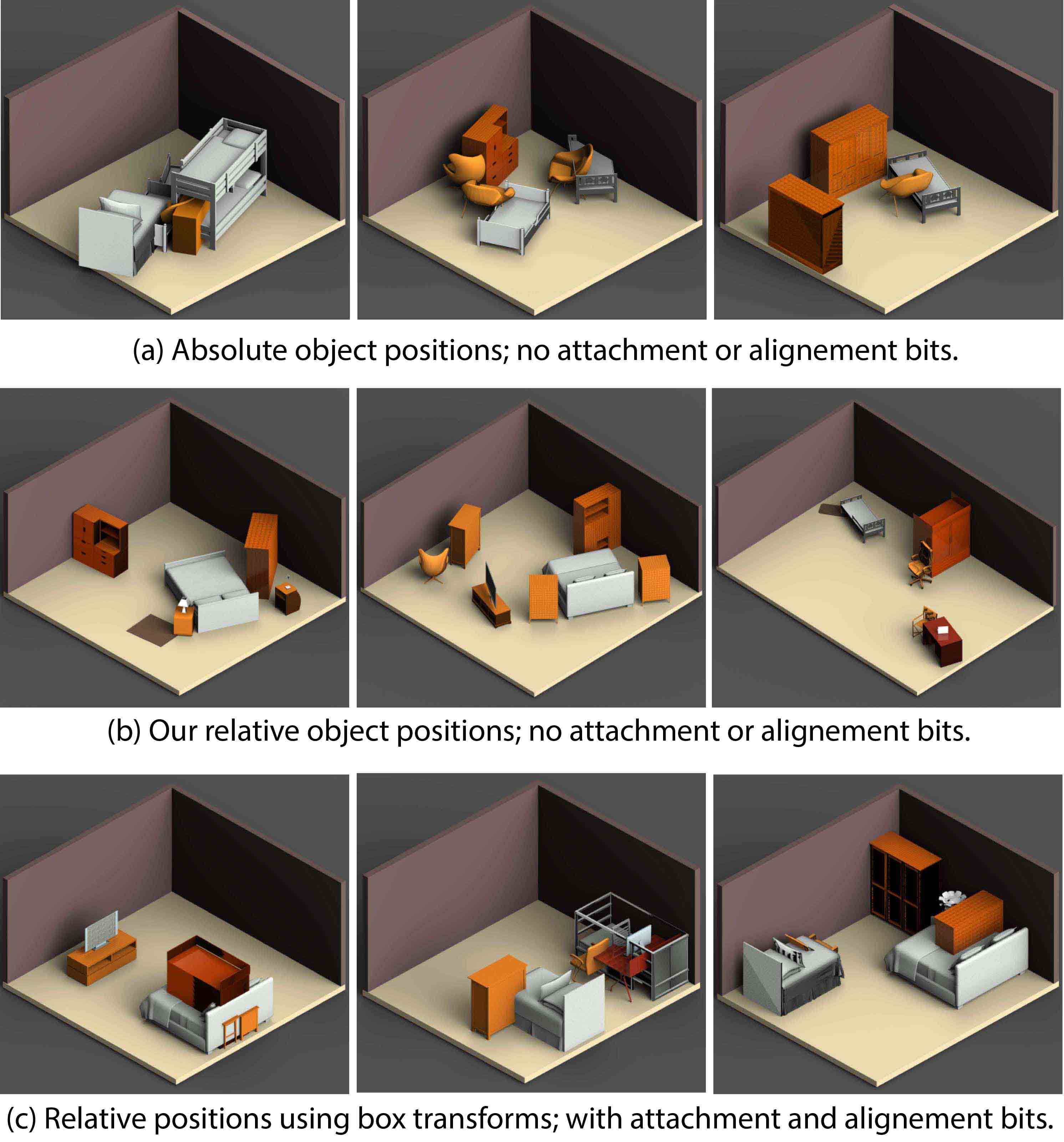}
	\vspace{-1em}
	\caption{Typical bedrooms generated using our method, but with alternate encodings of object position.}
	\vspace{-1em}
	\label{fig:alt-pos}
\end{figure}
\paragraph{Encoding of relative object positions}
With the key observation that predictabilities in object positions in a 3D scene are manifested in
{\em relative\/}, and not absolute, terms, we encode relative object positions using angles and offsets,
which are complemented with attachment and alignment bits to reinforce placement precisions.
To contrast this with the use of {\em absolute\/} object positions in the network, we show
in the first row of Figure~\ref{fig:alt-pos} several typical bedrooms generated with this latter option.
We can see that our RvNN-VAE was unable to learn plausible object placements: most objects are placed somewhat randomly without proper attachments or alignments, due to the lack of
predictability in their absolute positions.

In the second row of Figure~\ref{fig:alt-pos}, we show results of scene generation using our relative
position encoding but without alignment or attachment bits. As expected, while the rough relative
positioning of objects may be plausible, e.g., nightstands do surround the bed, precise alignments
and attachments, e.g., bed against the wall or nightstands against the bed, are not fulfilled.

In the final row of Figure~\ref{fig:alt-pos}, we show scenes generated using our RvNN-VAE, with relative
position encoding as described in Section~\ref{subsec:objpos}, except that the position of the target box
is encoded using a {\em single\/} translation vector between the centers of the target and reference
boxes. The results show that this alternative is problematic since box-to-box translations can vary
greatly, e.g., dressers can be placed in all directions around a bed. In such cases, the trained
network has the tendency to generate ``average'' translation vectors. This is why the dressers tend to
be placed near the center of the beds. In contrast, our representation which utilizes a binary indicator
vector to identify the closest edge pairs for offset encoding avoids the generation of average offset values.

\vspace{-0.5em}

\paragraph{Semantic labels}

\begin{figure}[!t]
	\centering
	\includegraphics[width=0.95\linewidth]{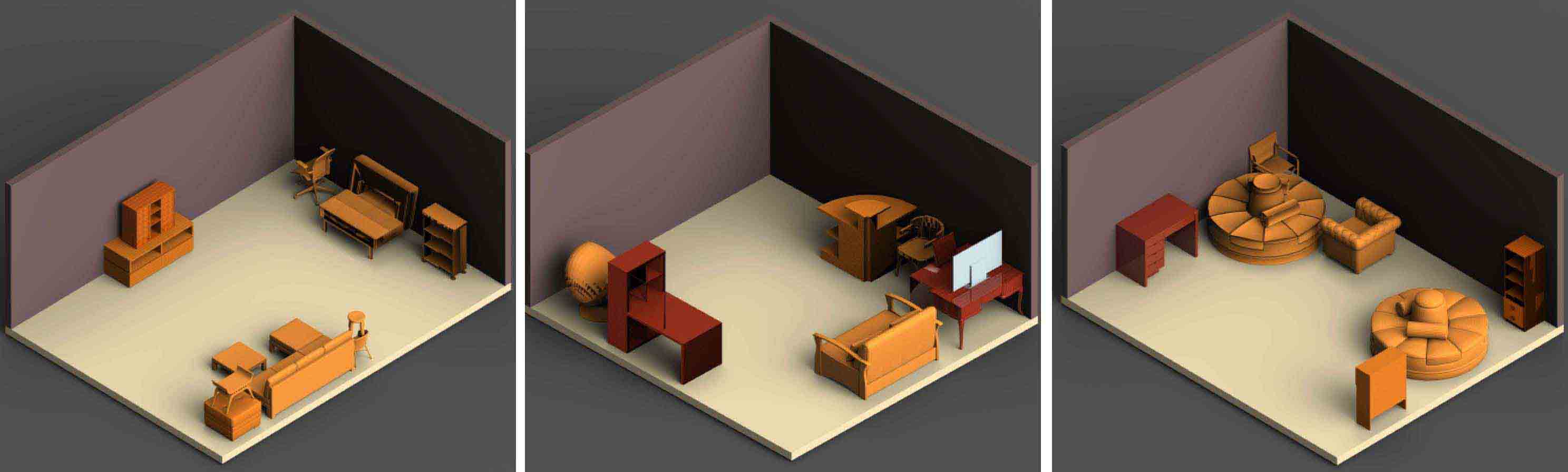}
	\caption{Typical bedroom scenes generated by our method when semantic labels are not included in object encodings in the leaf vectors, where 3D objects were retrieved based only on geometric information about the
		OBBs.}
	\label{fig:no_labels}
	\vspace{-1.5em}
\end{figure}
We incorporate semantic labels into the object encodings since object co-occurrences are
necessarily characterized by semantics, not purely geometric information such as OBB dimensions and
positions. As expected, removing semantic labels from the object encodings would make it difficult for
the network to learn proper object co-occurrences. As well, without semantic information associated with
the OBB's generated by the RvNN-VAE, object retrieval based on OBB geometries would also lead to clearly
implausible scenes, as shown by examples in Figure~\ref{fig:no_labels}.

\subsection{Applications}
\label{subsec:apps}
We present three representative applications of our generative model for 3D indoor scenes, highlighting its several advantages: 1) cross-modality scene generation;
2) generation of scene hierarchies; and 3) fast synthesis of large volumes of scene data.

\paragraph{2D layout guided 3D scene modeling.}

\begin{figure*}[!t]
	\includegraphics[width=0.9\textwidth]{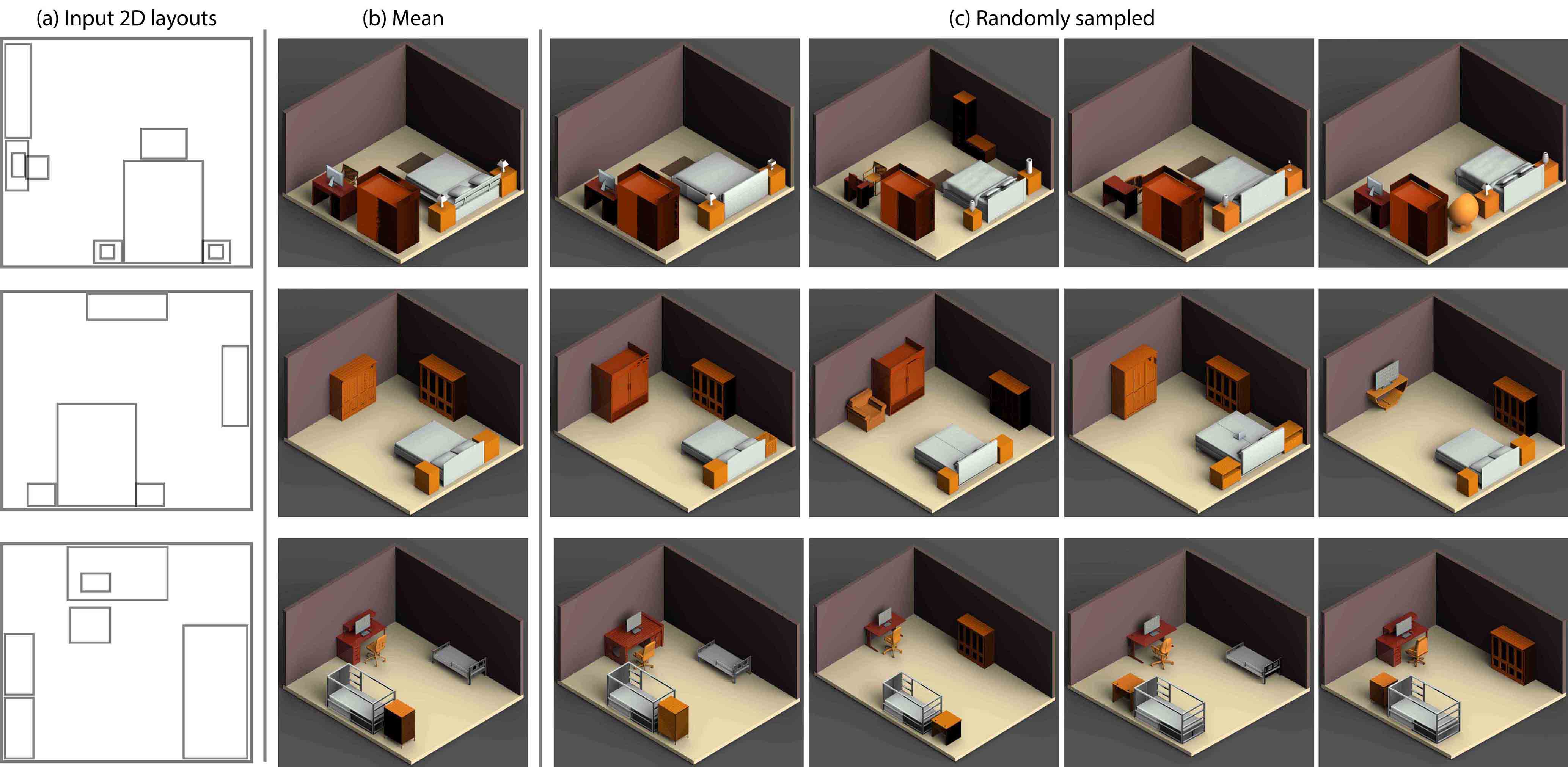}
	\caption{3D scene generation from 2D box layouts. Our trained RvNN-VAE can convert a 2D box layout representing the top-view of a scene (a) into a root code and then decode it into a 3D scene structure that closely resembles the 2D layout. (b) shows scenes decoded from the mean of the Gaussian of the latent vector and (c) shows scenes generated from latent vectors randomly sampled from the Gaussian.}
	\label{box2scene}
\end{figure*}

Several applications such as interior design heavily use 2D house plan, which is a top-view 2D box layout of an indoor scene. Automatically generating a 3D scene model from such 2D layout would be very useful in practice.
Creating a 3D scene from labeled boxes would be trivial.
Our goal is to generate a series of 3D scenes whose layout is close to the input boxes \textit{without} semantic labels, while ensuring a plausible composition and placement of objects.
To do so, we encode each input 2D box into a leaf vector with unknown class (uniform probability for all object classes). We then construct a hierarchy based on the leaves, obtaining a root code encoding the entire 2D layout.
Note that the \textit{support} relation changes to \textit{overlap} in 2D layout encoding.
The root code is then mapped into a Gaussian based on the learned VAE.
A sample from the Gaussian can be decoded into a hierarchy of 3D boxes \emph{with labels} by the decoder, resulting in a 3D scene.
%
Thus, a set of scenes can be generated, whose spatial layouts closely resemble the input 2D layout.
A noteworthy fact is that both the encoder and decoder used here are pretrained on 3D scenes; no extra training is required for this 2D-to-3D generation task. This is because 2D box layout is equivalent to 3D one in terms of representing spatial layout and support relations, which makes the learned network reusable for cross-modality generation.

Figure \ref{box2scene} shows a few examples of 2D layout guided 3D scene generation, where no label information was used. The missing label information is recovered in the generated scenes. 
In the generated 3D scenes, many common sub-scenes observed from the training dataset are preserved while fitting to the input 2D layout. 

\paragraph{Hierarchy-guided scene editing}

\begin{figure*}[!t]
	\includegraphics[width=\textwidth]{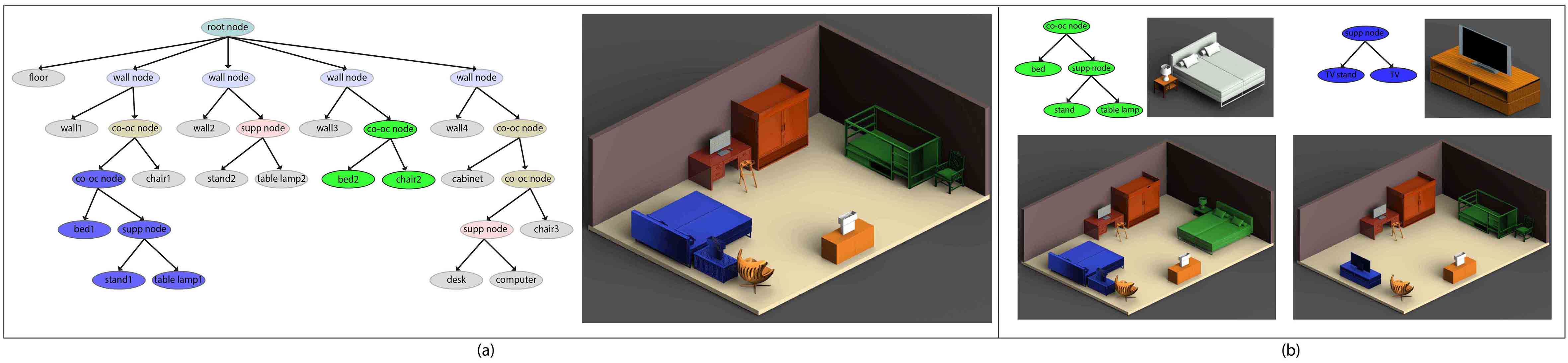}
	\caption{Hierarchy-guided scene editing. Given a generated scene and its hierarchy (a), user can select a subtree (colored in green or blue) corresponding to a sub-scene (objects colored similarly). The subtree (sub-scene) can be replaced by another from other generated hierarchies, resulting in an edited scene (b).}
	\label{editing_hierarchy}
\end{figure*}
During scene modeling, it is desirable for the users to edit the generated scenes to reflect their intent. With the hierarchies accompanied by the generated scenes,
our method enables the user to easily edit a scene at different granularities in an organized manner.
Both object level and sub-scene level edits are supported. 
Specifically, the users can select a node or a group of nodes in the hierarchy to alter the scene. Once the node(s) is selected, the user can simply delete or replace the subtree of the selected nodes, or move them spatially, as shown in Figure \ref{editing_hierarchy}. To replace a subtree, colored in Figure \ref{editing_hierarchy}(a), the user can simply reuse a subtree chosen from another automatically generated hierarchy; see Figure \ref{editing_hierarchy}(b). To prepare candidate subtrees used to replace the selected one, we retrieve, from other hierarchies, the subtrees whose sibling subtrees are similar to that of the selected one.

\paragraph{Data enhancement for training deep models}
Efficient generation of large volumes of 
diverse 3D indoor scenes provides a means of data enhancement for training deep models, potentially boosting their performance for scene understanding tasks. To test this argument, we train PointNet \cite{qi2017pointnet}, which was originally designed for shape classification and segmentation, for the task of semantic scene segmentation. Using the same network design and other settings as in the original work, we train PointNet on scenes both from SUNCG dataset and those generated by our method. To adapt to PointNet training, each scene is sliced into $16$ patches containing $4,096$ points, similar to \cite{qi2017pointnet}. For each experiment, we randomly select the training and testing data (non-overlapping) and report the results.

\begin{table}
	\begin{tabular} {| c | c | c |}
		\hline
		\textbf{TRAINING DATA}& \textbf{TEST DATA}& \textbf{Accuracy} \\
        \hline
        {SUNCG, 12K} & {SUNCG, 4K} & {76.29\%} \\
		{SUNCG, 12K} & {Ours, 4K} & {41.21\%} \\
		{SUNCG, 12K} & {Ours, 2K + SUNCG, 2K} & {60.13\%} \\
		\hline
        {Ours, 12K} & {SUNCG, 4K} & {58.04\%} \\
        {Ours, 12K} & {Ours, 4K} & {77.03\%} \\
        {Ours, 12K} & {Ours, 2K + SUNCG, 2K} & {69.86\%}  \\
        \hline
        {SUNCG, 6K + Ours, 6K} & {SUNCG, 4K} & {70.19\%} \\
        {SUNCG, 6K + Ours, 6K} & {Ours, 4K} & {77.57\%} \\
        {SUNCG, 6K + Ours, 6K} & {Ours, 2K + SUNCG, 2K} & {77.10\%} \\
        \hline
        {SUNCG, 12K + Ours, 12K} & {SUNCG, 8K} & {74.45\%} \\
        {SUNCG, 12K + Ours, 12K} & {Ours, 8K} & {81.95\%} \\
        {SUNCG, 12K + Ours, 12K} & {Ours, 4K + SUNCG, 4K} & {80.97\%} \\
		\hline
	\end{tabular}
\vspace{3pt}
	\caption{Performance of semantic scene segmentation using PointNet trained and tested on different datasets. Our generated scenes can be used to augment SUNCG to generalize better.}
	\label{PointNet_exp}
\vspace{-15pt}
\end{table}


Table \ref{PointNet_exp} presents a mix of experiments performed on SUNCG dataset and our generated scenes.
Training and testing on pure SUNCG gets a higher performance than testing on "Ours" or the mixture of data. The same is true when we train and test on "Ours". This means that training purely on either of the datasets doesn't generalize well and leads to model overfitting, as seen in the first two rows. Note that training and testing on pure SUNCG gets the highest performance among all the experiments tested on SUNCG data. It's because the former model was overfitting on SUNCG.
However, when we 
test on the mixed data (SUNCG + Ours), 
the network's ability to generalize better is improved when the two datasets are combined for training. In other words, the network becomes robust to diversity in the scenes. 
Thus, instead of \textit{randomly} enhancing the scene data, one can use our generated scenes (which are novel, diverse and plausible) as a reserve set for data enhancement to efficiently train deep models.

\section{Discussion, Limitations, and Future Work}
\label{sec:future}

Our work makes a first attempt at developing a generative neural network to learn hierarchical structures of 3D indoor scenes. The network, coined GRAINS, integrates a recursive neural network with a variational autoencoder, enabling us to generate a novel, plausible 3D scene from a random vector in less than a second. The network design consists of several unique elements catered to learning scene structures for rectangular rooms, e.g., relative object positioning, encoding of object semantics, and use of wall objects as initial references, distinguishing itself from previous generative networks for 3D shapes~\cite{wu2016,li2017}.


As shown in Section~\ref{subsec:justify}, the aforementioned design choices in GRAINS are responsible for improving the plausibility of the generated scenes. However, one could also argue that they are introducing certain ``handcrafting'' into the scene encoding and learning framework.
It would have been ideal if the RvNN could: a) learn the various types of object-object relations on its own from the data instead of being designed with three pre-defined grouping operations; b) rely only on the extracted relations to infer all the object semantics, instead of encoding them explicitly; and c) work with a more generic representation of object positions than our current special-purpose encoding of relative positionings. At this point, these are all tall challenges to overcome. An end-to-end, fully convolutional network that can work at the voxel level while being capable of generating clean and plausible scene structures also remains elusive.

Another major limitation of GRAINS, and of generative recursive autoencoders such as GRASS~\cite{li2017}, is that we do not have direct control over objects in the generated scene. For example, we cannot specify object counts or constrain the scene to contain a subset of objects. Also, since our generative network produces only labeled OBBs and we rely on a fairly rudimentary scheme to retrieve 3D objects to fill the scene, there is no precise control over the shapes of the scene objects or any fine-grained scene semantics, e.g., style compatibility. It is possible to employ a suggestive interface to allow the user to select the final 3D objects. Constrained scene generation, e.g., by taking as input a partial scene or hierarchy, is also an interesting future problem to investigate.


Modeling indoor scene structures in a hierarchical way does hold its merits, as demonstrated in this work; it is also a natural fit to RvNNs. However, hierarchies can also be limiting, compared to a more flexible graph representation of scene object relations. For example, it is unclear whether there is always a sensible hierarchical organization among objects which may populate a messy office desk including computer equipments, piles of books, and other office essentials. Also, in many instances, it can be debatable what the best hierarchy for a given scene is. To this end, recent works on learning generative models of graphs may be worth looking into.

Finally, as a data-driven method, GRAINS can certainly produce unnatural object placements and orientations. A few notable failure cases can be observed in Figure~\ref{fig:gallery}, e.g., a computer display and a swivel chair facing the wall, shown in the fifth row of column (a). One obvious reason is that the training data is far from perfect. For example, object placements in some of the SUNCG scenes can be unnatural, as shown in the last row of column (b), where a bed is placed in the middle of the room. More critically, the training scene hierarchies in our work were produced heuristically based on spatial object relations and as such, there is no assurance of consistency even among similar scenes belonging to the same scene category. In general, computing consistent hierarchies over a diverse set of scenes is far from straightforward. Comparatively, annotations for pairwise or group-wise object relations are more reliable. An intriguing problem for future work is to develop a generative network, possibly still based on RvNNs and autoencoders, that are built on learning {\em subscene structures\/} rather than global scene hierarchies.


\begin{acks}
	We thank the anonymous reviewers for their valuable comments. This work was supported, in parts, 
	by an NSERC grant (611370), the 973 Program of China under Grant 2015CB352502, key program of NSFC (61332015), NSFC programs (61772318, 61532003, 61572507, 61622212), ISF grant 2366/16 and gift funds from Adobe. Manyi Li was supported by the China Scholarship Council.
\end{acks}

\bibliographystyle{ACM-Reference-Format}
\bibliography{scenes}

\pagebreak
\setcounter{section}{0}
\setcounter{figure}{0}
\setcounter{table}{0}
\setcounter{page}{1}
\begin{titlepage}
	\vspace*{\stretch{1.0}}
	\begin{center}
		\Large\textbf{Supplemental Material \newline GRAINS: Generative Recursive Autoencoders for INdoor Scenes}\\
	\end{center}
	\vspace*{\stretch{1.0}}
\end{titlepage}

\section{More scene generation results}
\begin{figure*}
	\centering
	\subfloat{\includegraphics[width=\textwidth]{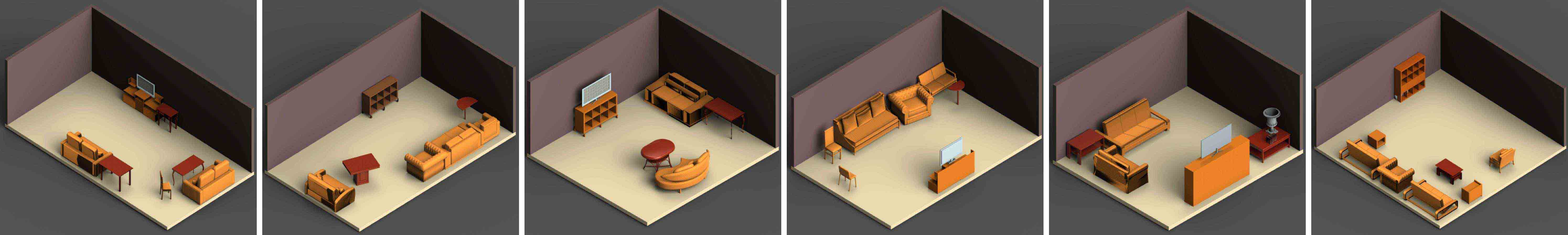}}
	\qquad
	\subfloat{\includegraphics[width=\textwidth]{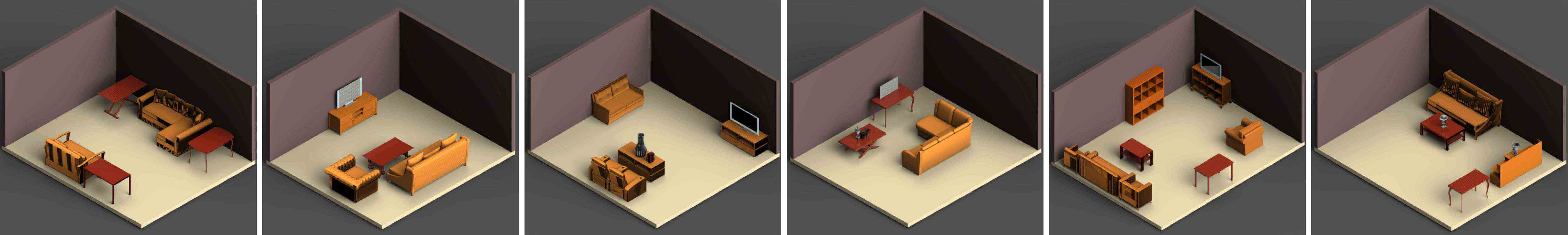}}
	\caption*{(a) Generated living rooms}
	\subfloat{\includegraphics[width=\textwidth]{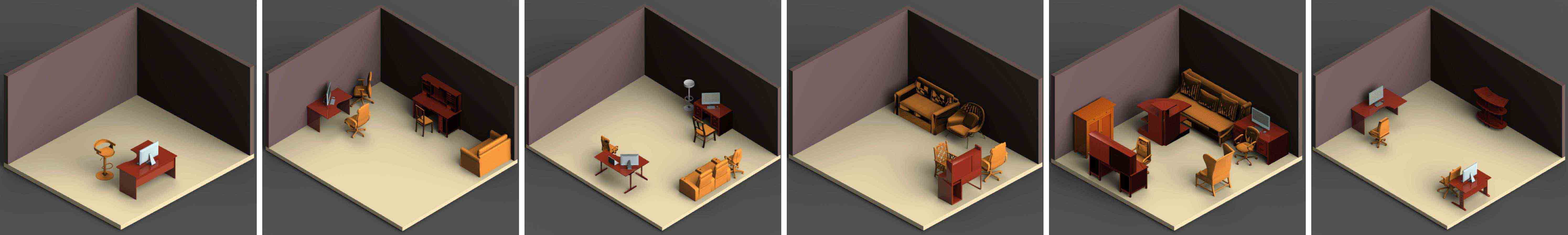}}
	\qquad
	\subfloat{\includegraphics[width=\textwidth]{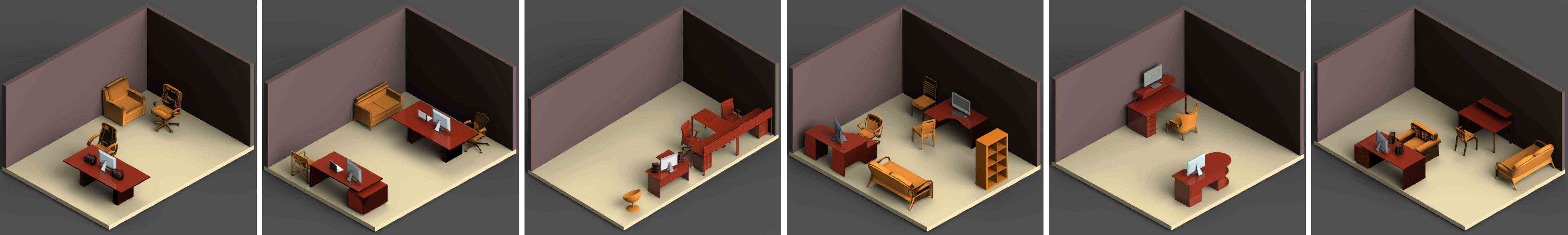}}
	\caption*{(b) Generated offices}
	\subfloat{\includegraphics[width=\textwidth]{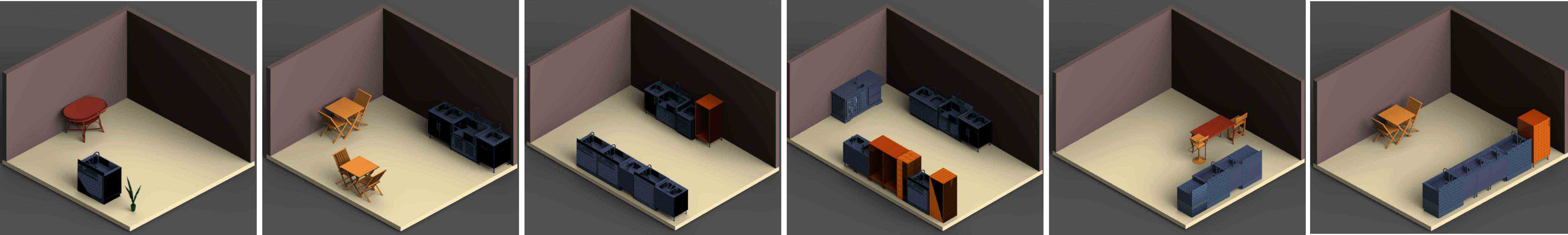}}
	\qquad
	\subfloat{\includegraphics[width=\textwidth]{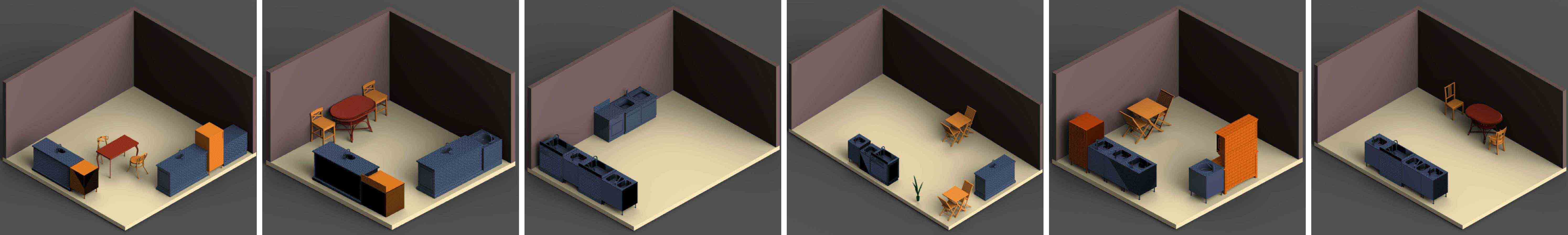}}
	\caption*{(c) Generated kitchens}
	\caption{More results of other categories. Top two rows: generated living rooms; middle two rows: generated offices; last two rows: generated kitchens.}
	\label{moreresults}
\end{figure*}

We present more scene generation results for living rooms, kitchens, and offices in figure \ref{moreresults}. We train a separate RvNN-VAE for each room category, using the same settings as mentioned in the main paper. Our network learns the common patterns present in various sub-scenes, in each room category; for example:(a) sofa and table in living rooms, (b) desk, computer and chair in offices, and (c) set of \textit{attached} kitchen cabinets in kitchens. Note that for kitchens, the training set is biased towards kitchen-cabinets, which limits the variety of object classes in the generated scenes. The key characteristic, however, is the attachment of kitchen-cabinets, which are reflected ably by our generated scenes. Our RvNN-VAE framework not only learns to preserve the object patterns in the sub-scenes (thus making the generated scenes plausible), but also enables diversity by incorporating multiple sub-scenes into a room and placing them accordingly.

\section{Object co-occurrence}
\begin{figure*}
	\centering
	\includegraphics[width=0.24\textwidth]{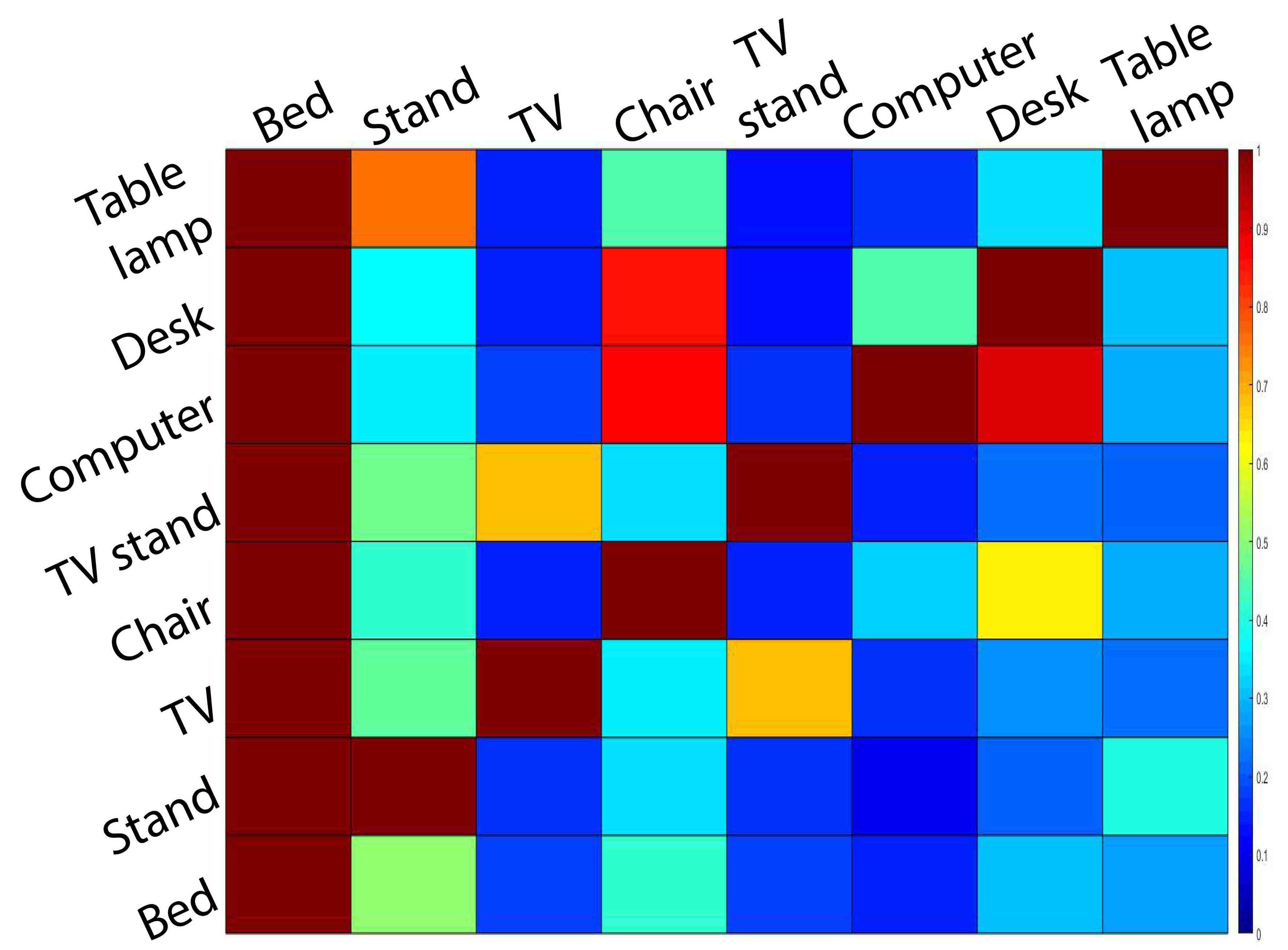}
	\includegraphics[width=0.24\textwidth]{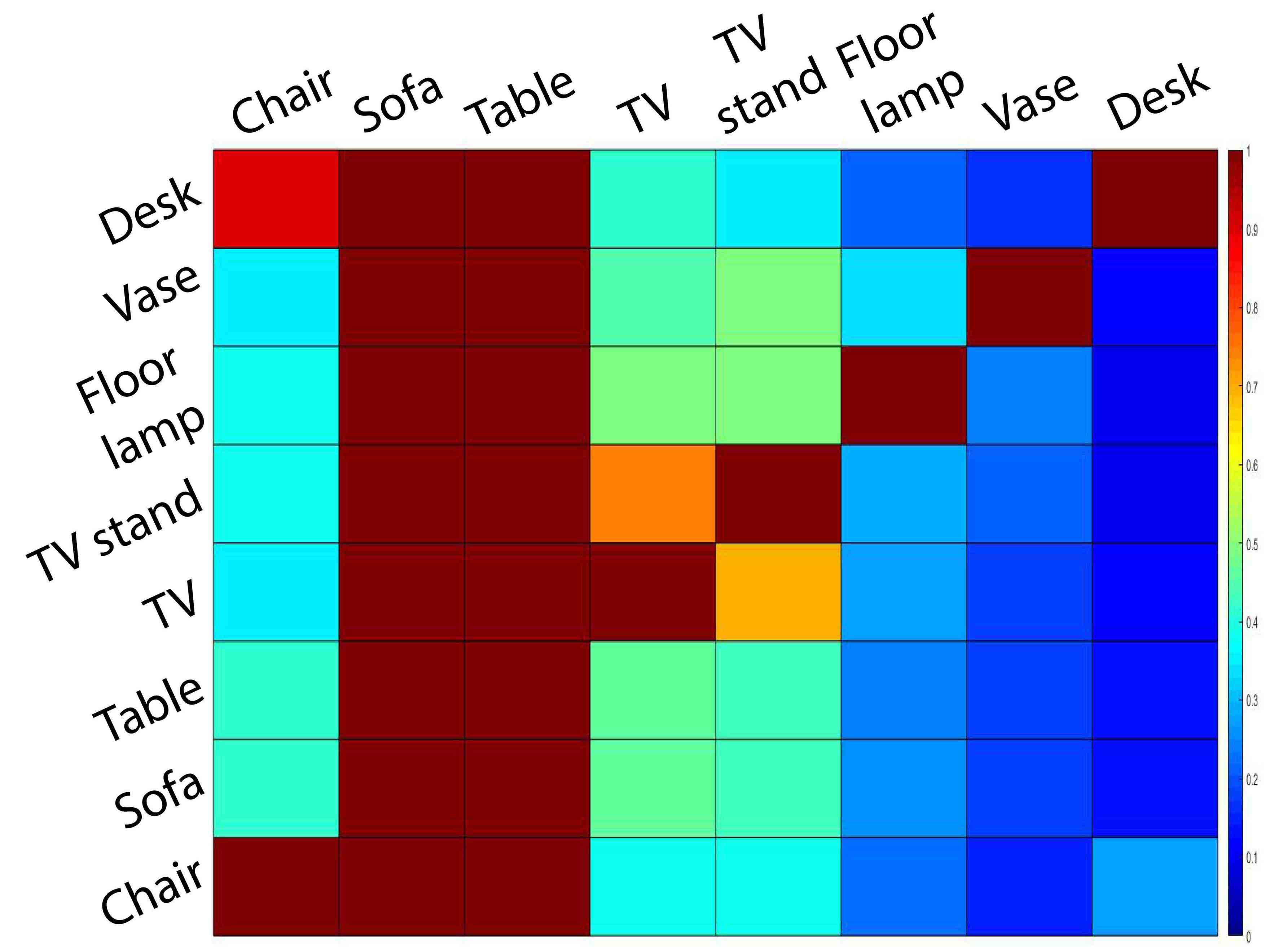}
	\includegraphics[width=0.24\textwidth]{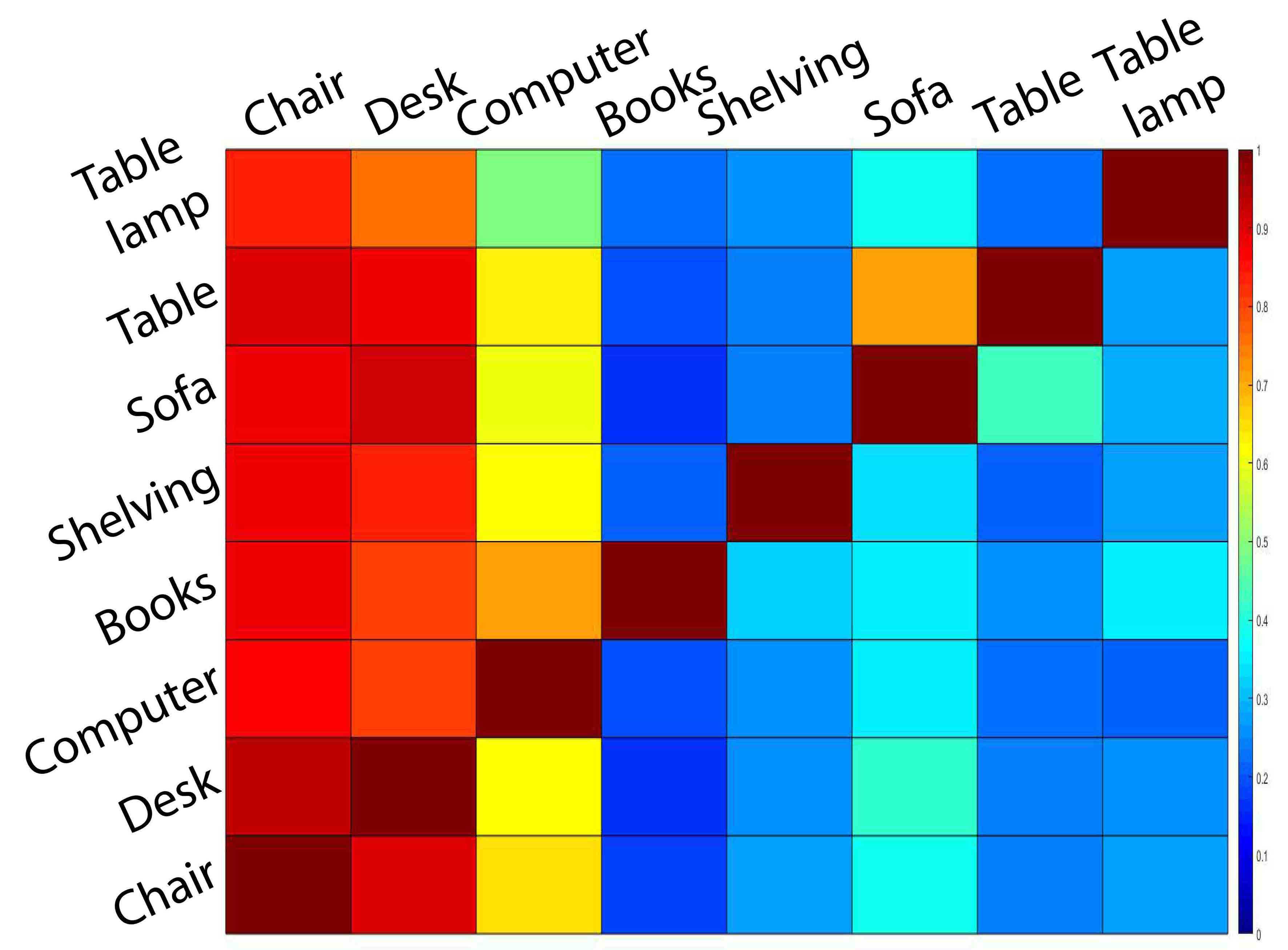}
	\includegraphics[width=0.24\textwidth]{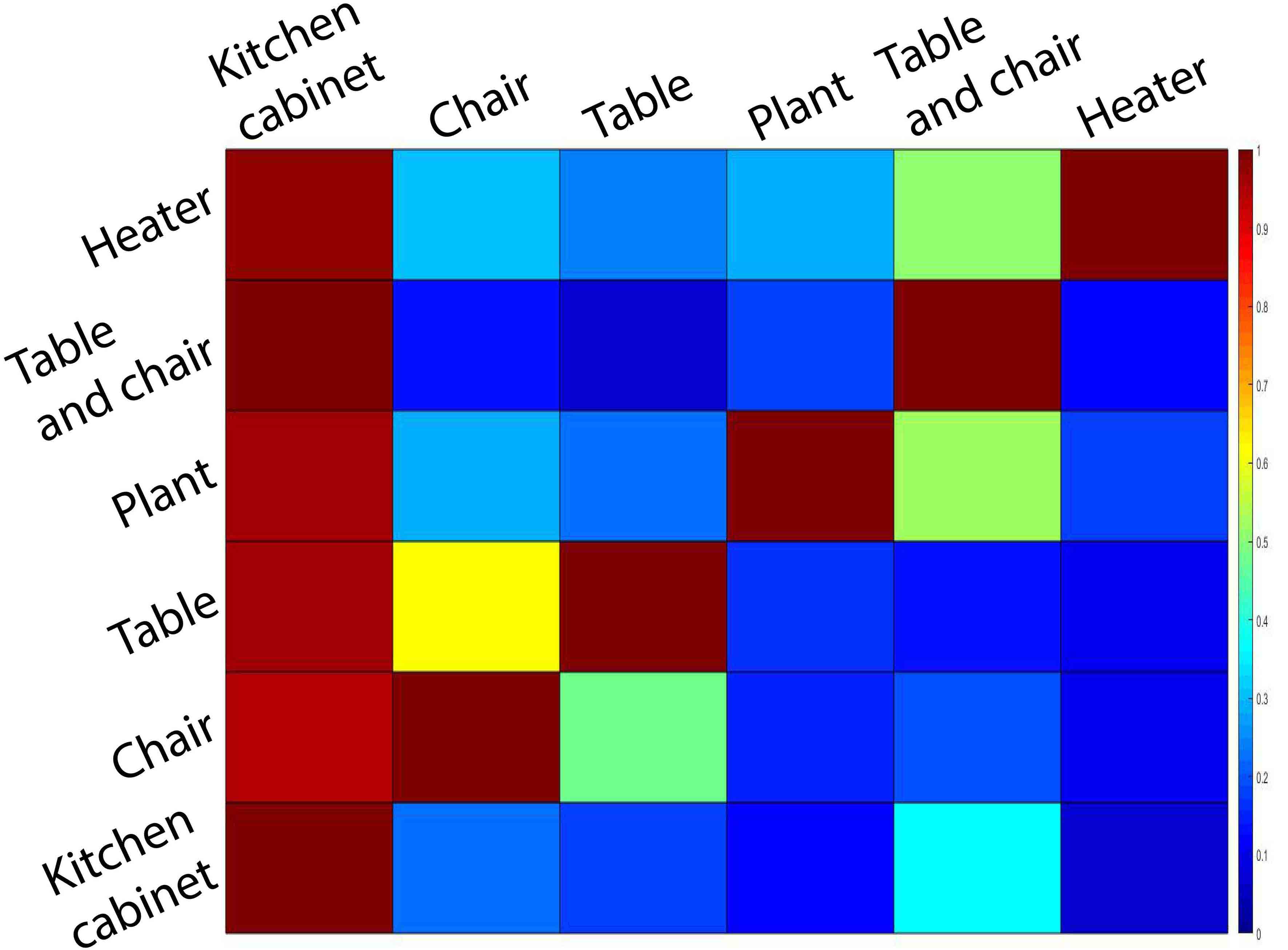}
	\includegraphics[width=0.24\textwidth]{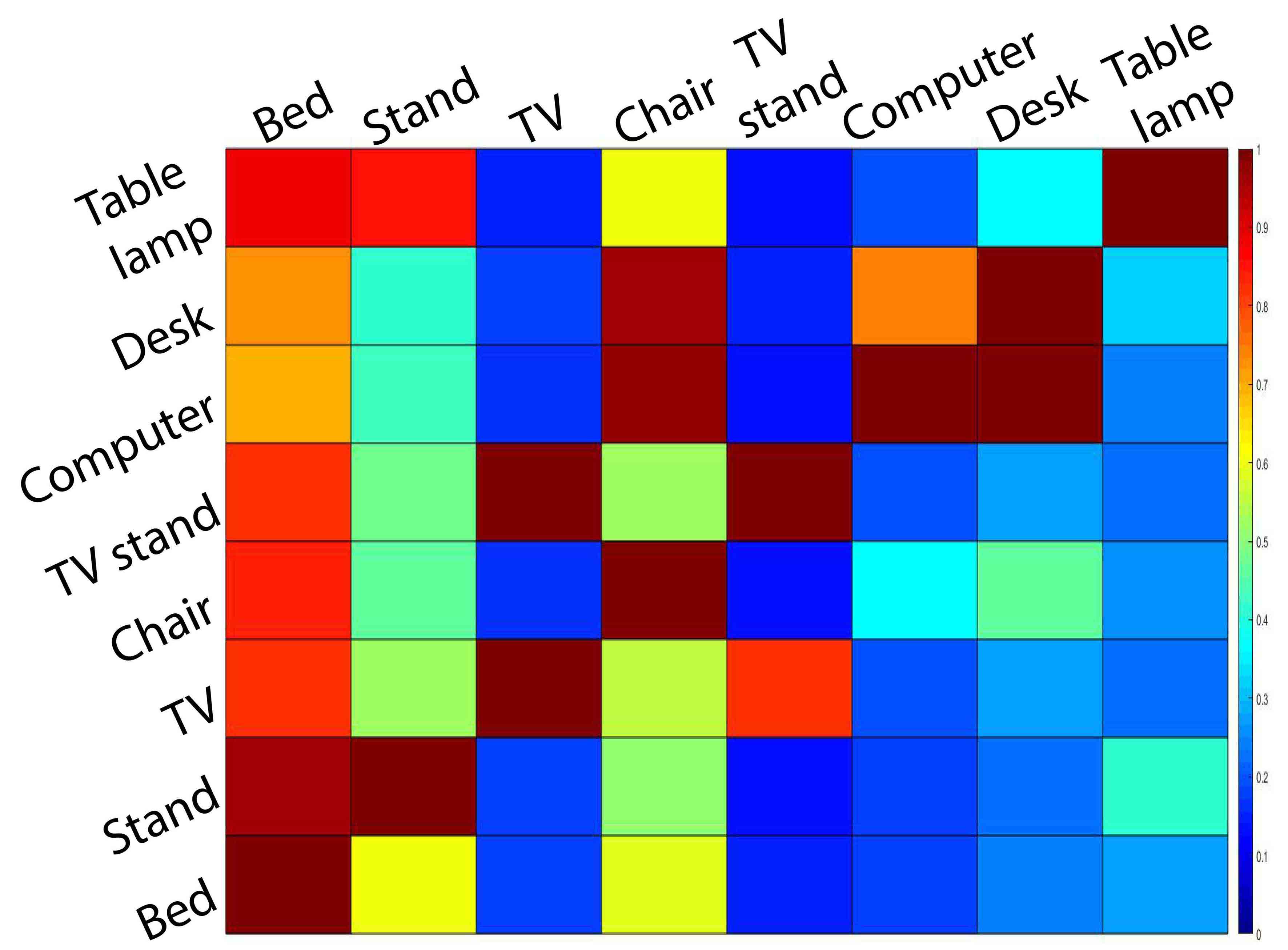}
	\includegraphics[width=0.24\textwidth]{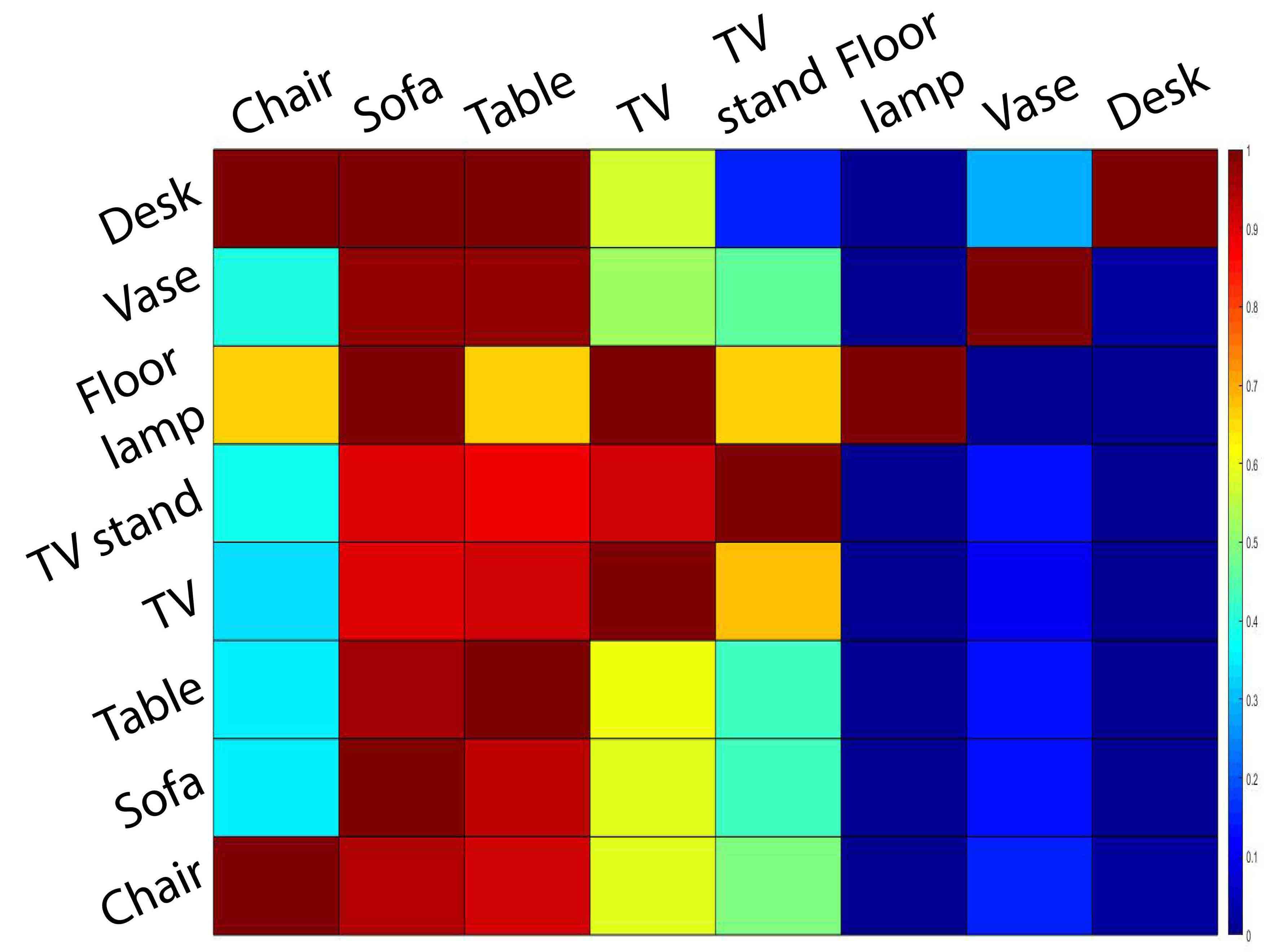}
	\includegraphics[width=0.24\textwidth]{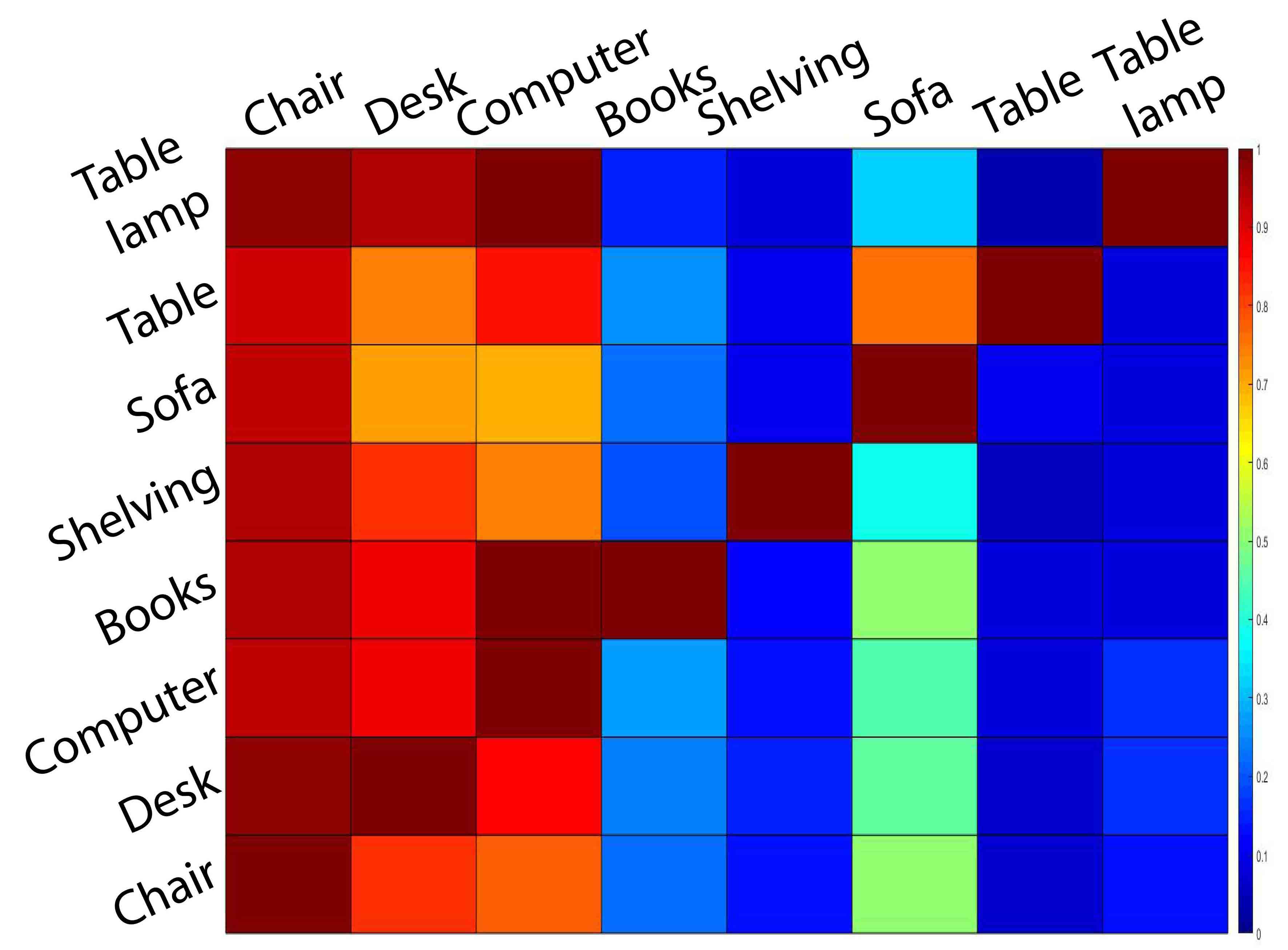}
	\includegraphics[width=0.24\textwidth]{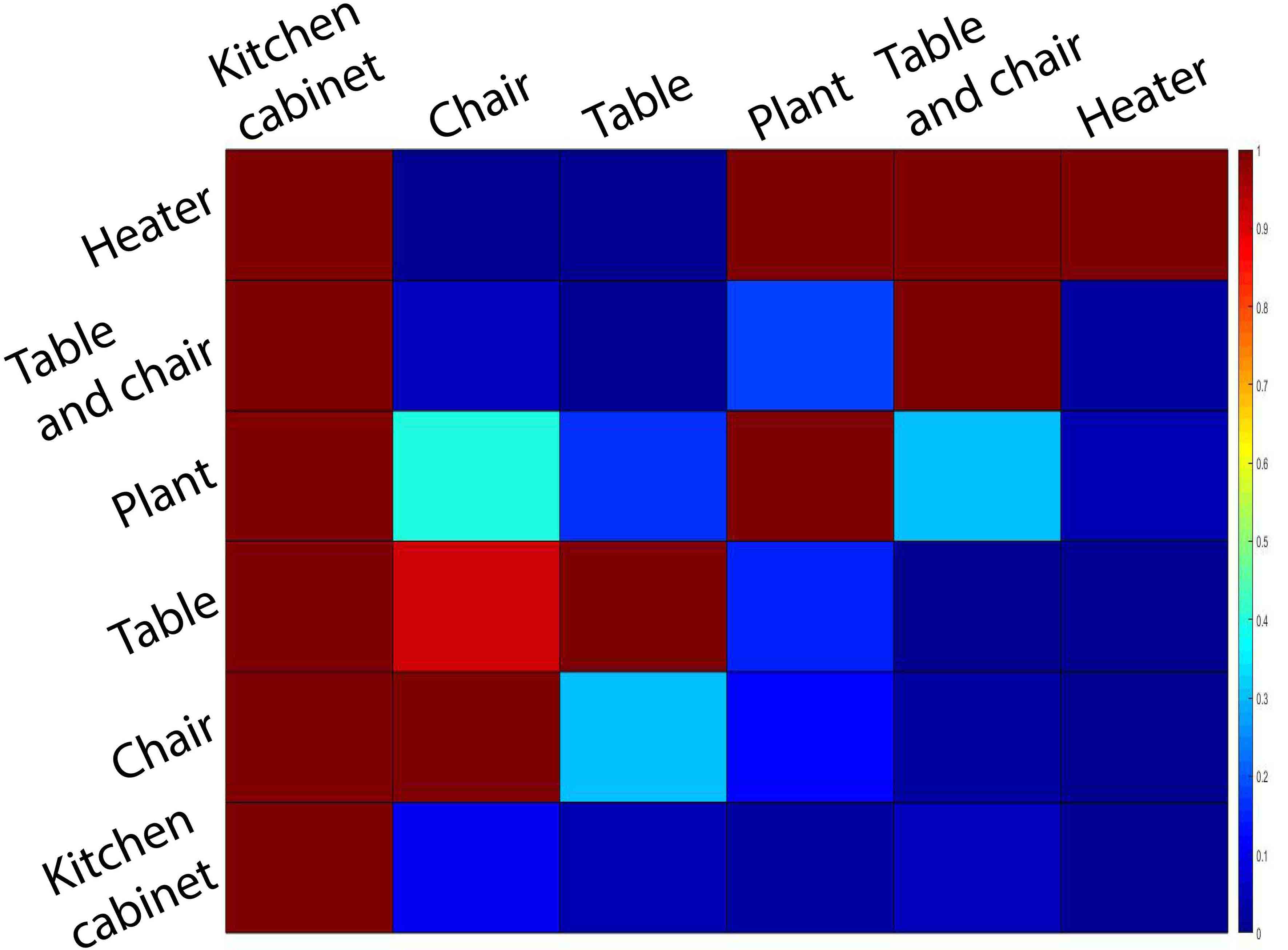}
	\caption{The object co-occurrence in the training scenes (top row) and our generated scenes (bottom row). From left to right are the plots of bedroom, living room, office, kitchen. Higher probability values correspond to warmer colors. Note that we only plot the co-occurrence between objects belonging to the frequent classes.}
	\label{probs}
\end{figure*}

\begin{figure*}
	\centering
	\includegraphics[width=0.24\textwidth]{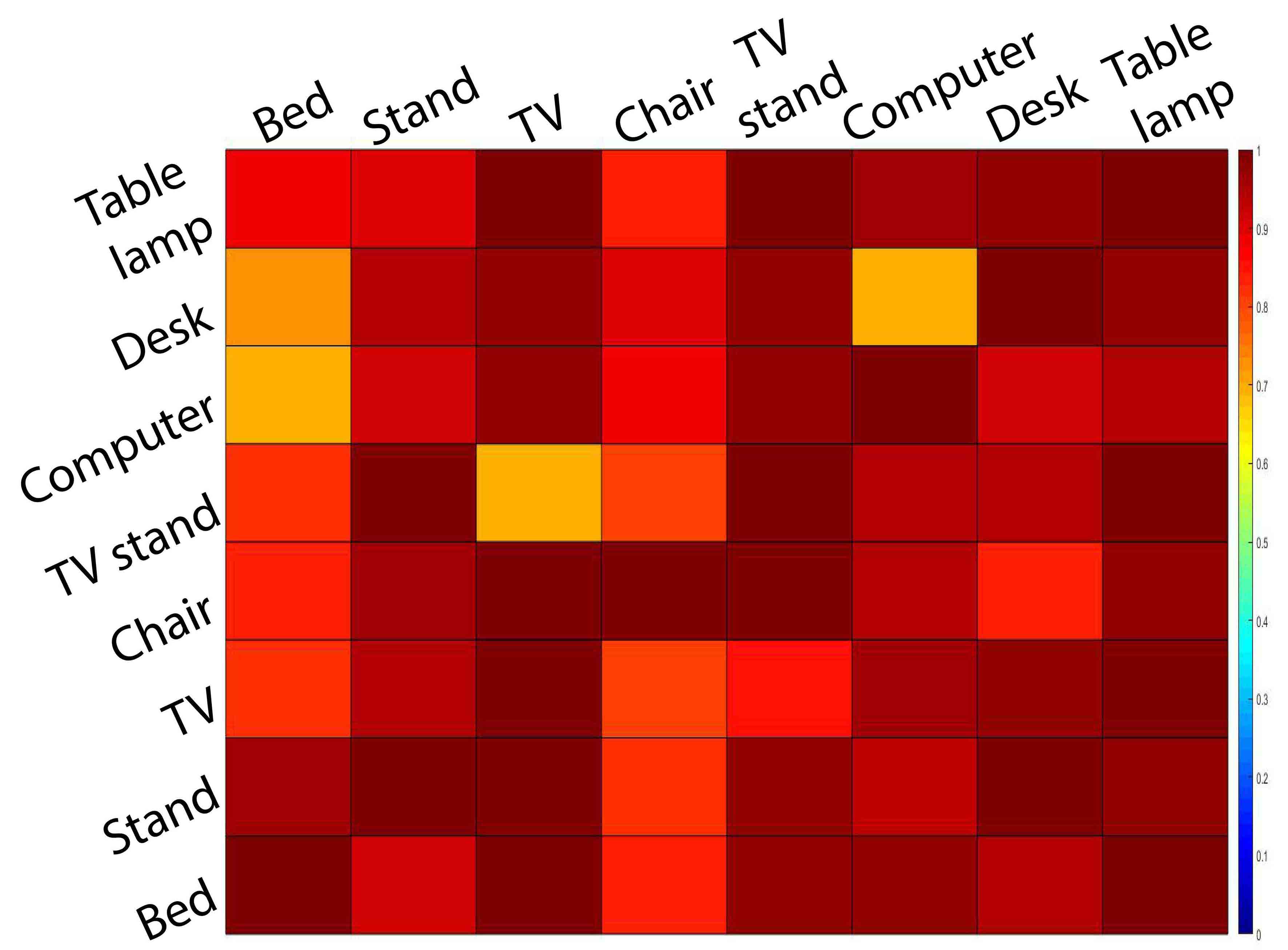}
	\includegraphics[width=0.24\textwidth]{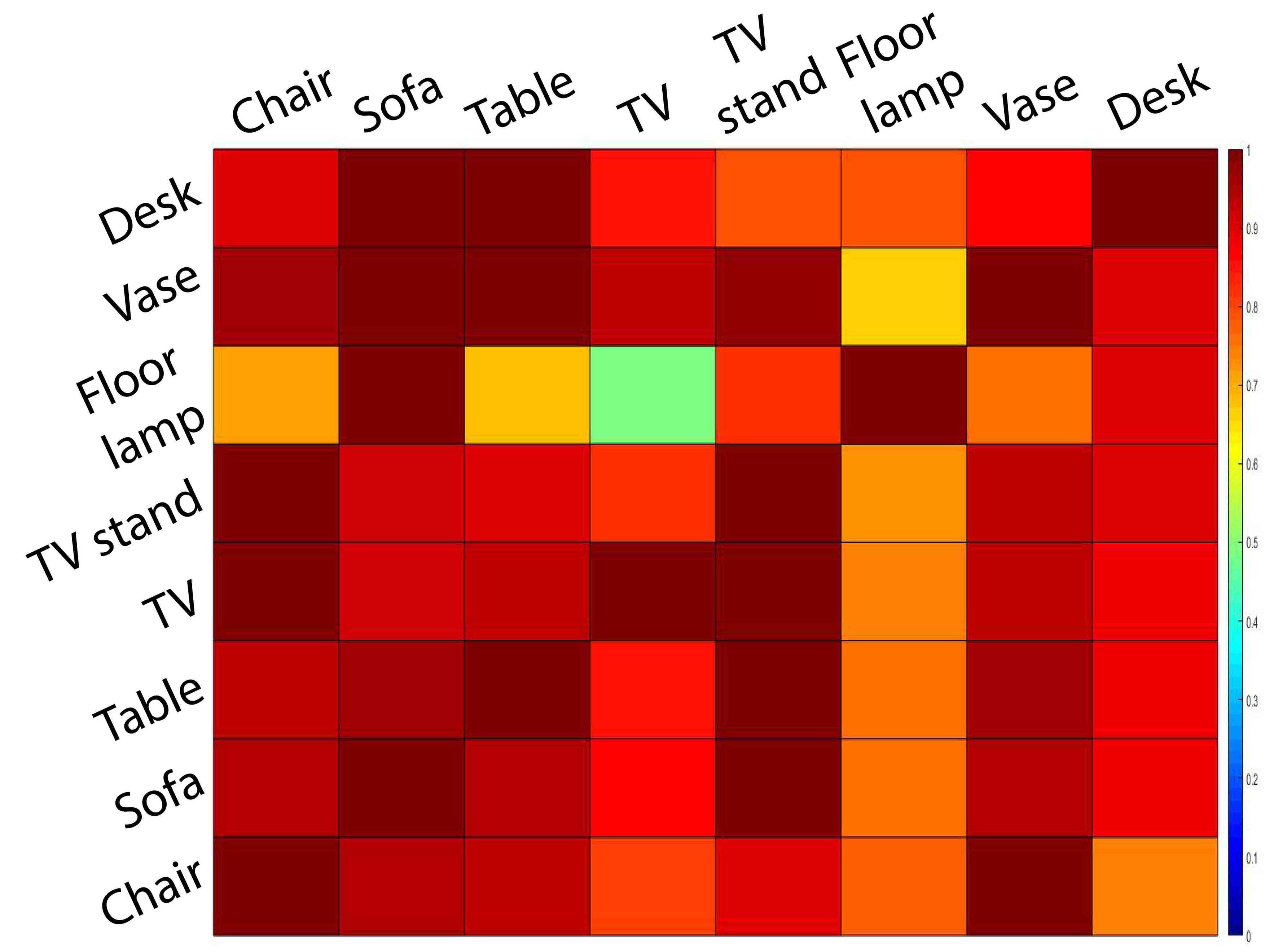}
	\includegraphics[width=0.24\textwidth]{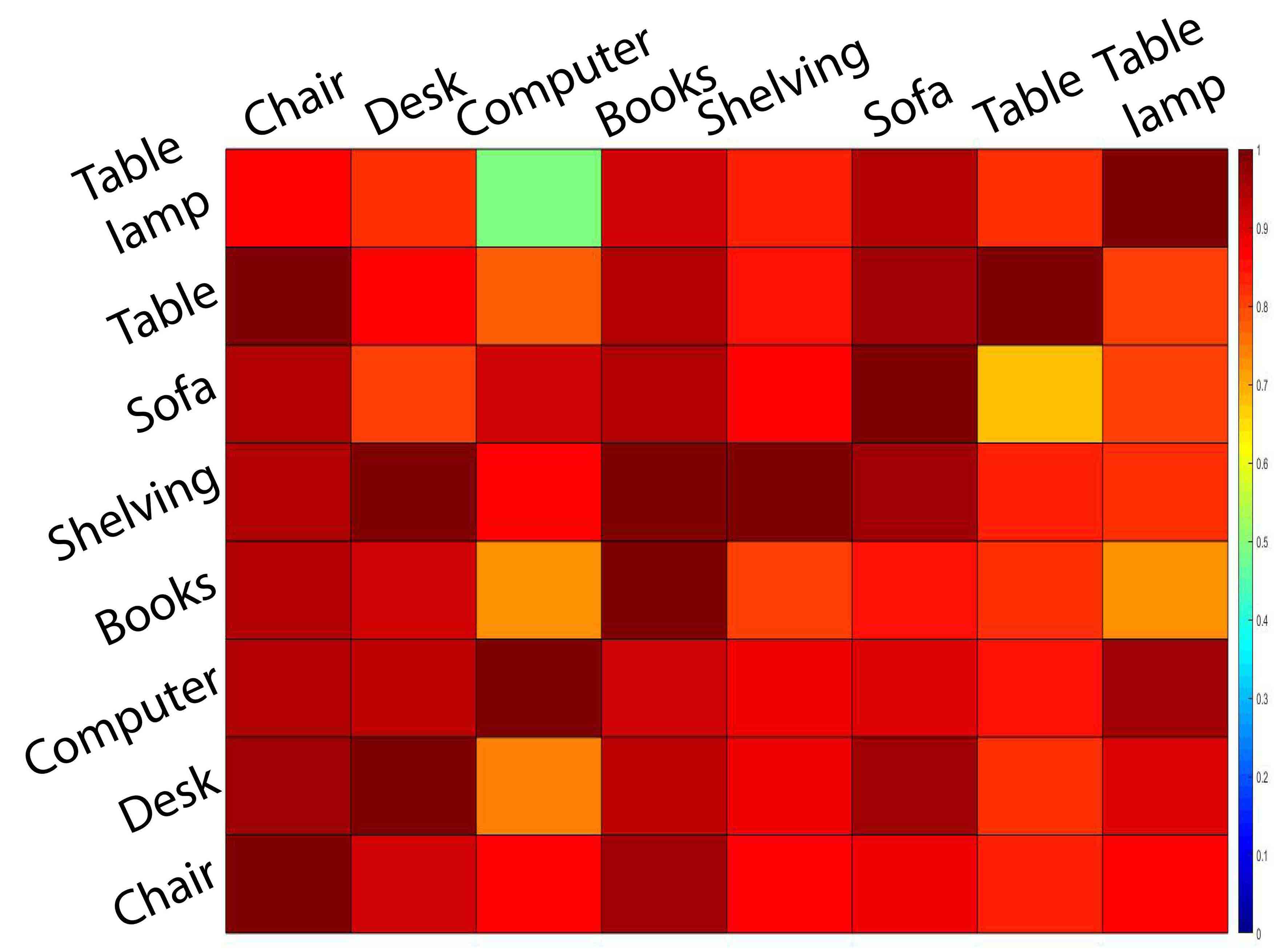}
	\includegraphics[width=0.24\textwidth]{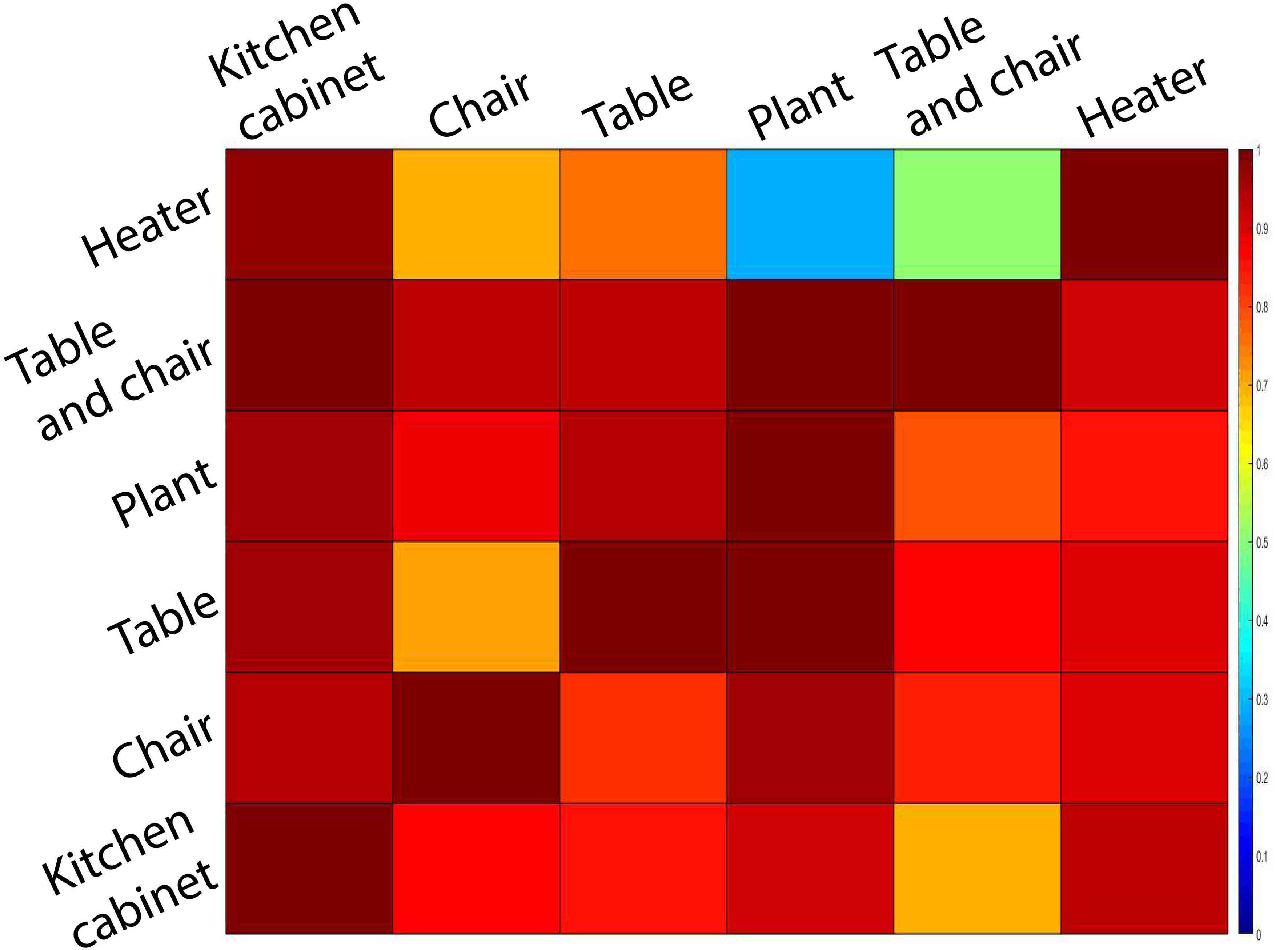}
	\caption{Similarity plots between object co-occurrence in the training set vs. in the RvNN-generated scenes. Left to right: bedroom, living room, office and kitchen.}
	\label{sims}
\end{figure*}

One possible way to validate our learning framework is to examine its ability to reproduce the probabilities of object co-occurrence in the training scenes. To measure the co-occurrence, we define a conditional probability $P(c_1|c_2)=\frac{N(c_1,c_2)}{N(c_2)}$, where $N(c_1,c_2)$ is the number of scenes with \textit{at least} one object belonging to category $c_1$ and one object belonging to category $c_2$. Figure \ref{probs} plots the object co-occurrences in the training scenes (top row) and our generated scenes (bottom row). Figure \ref{sims} is a similarity plots between the training set and the RvNN-VAE generated results $s(c_1|c_2)=1-|P_t(c_1|c_2)-P_g(c_1|c_2)|$. After training, our network is able to generate scenes with similar object occurrences compared to the training scenes.

\section{Distribution of the relative positions}
\begin{figure*}
	\centering
	\includegraphics[width=\textwidth]{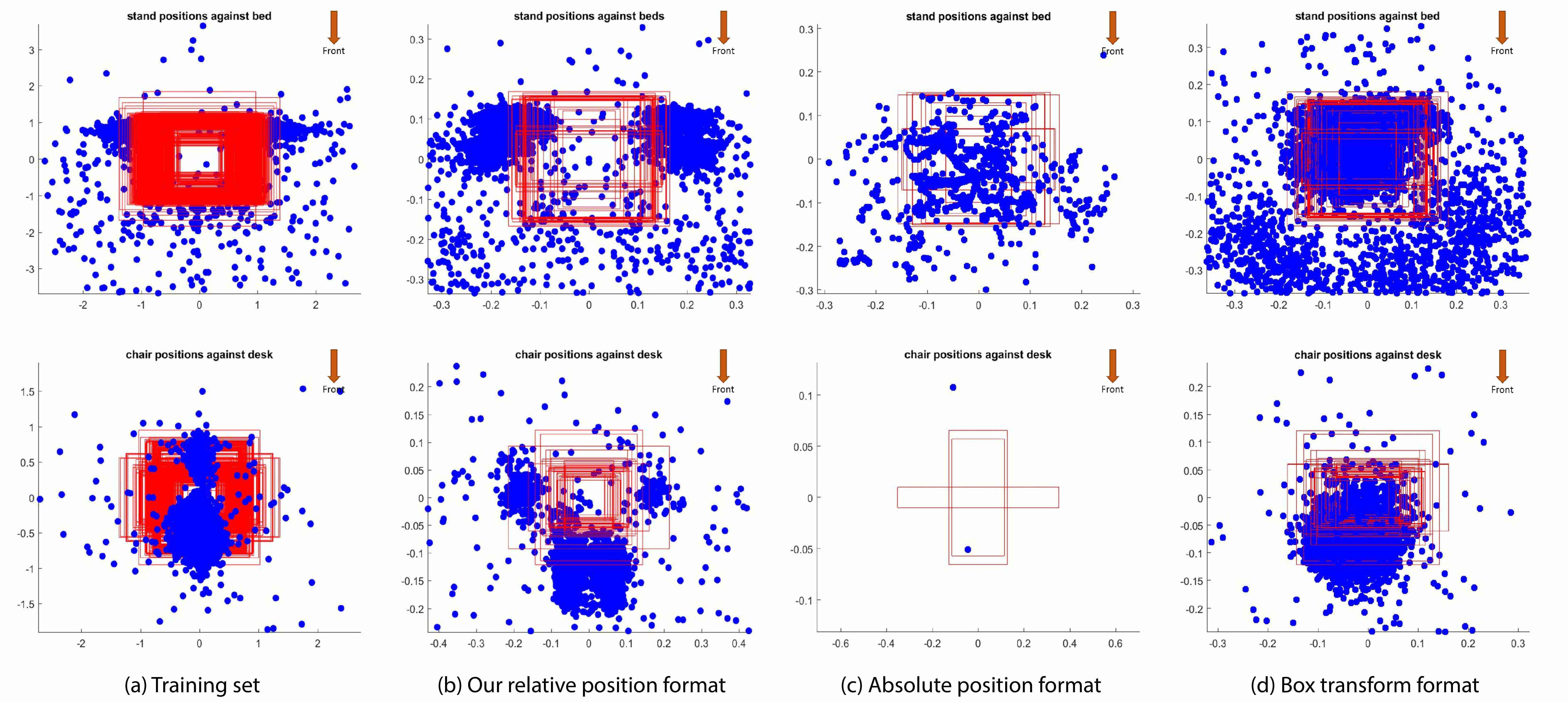}
	\caption{The relative position distribution. The relative positions are obtained from: (a) indoor scenes from the training set; generated scenes each using (b) our relative position format, (c) absolute position format and (d) box transforms, respectively. The first row shows the center points of stands (blue dots) and beds (red rectangles), under the local frames of the beds. The second row shows the center points of chairs (blue dots) and desks(red rectangles), under the local frames of the desks. The arrows show the front orientations of beds and desks. Best viewed in color on Adobe Reader.}
	\label{distribution_pos}
\end{figure*}

To further support our learning framework, we observe the distribution of relative positions between relevant objects. Conforming to common positioning patterns would reflect a good framework. Figure \ref{distribution_pos} shows the distributions of relative positions between object categories, from the training set (first column) and scenes generated using different relative position formats(ours, absolute position and box transforms). The first row shows the distribution of the relative positions of nightstands against the nearest bed. The blue dots represent the center points of the nightstands, and the red boxes represent the beds. The orientations of the beds are given by the arrow on the top right corner. For the bed-nightstand pair, our method outperforms the other two relative positions as it learns to put most of the nightstands on the left and/or right side of the beds. Consider the chair-desk pair  shown in the second row of figure \ref{distribution_pos}, the training set has most chairs in front of the desks and also has some chairs in the back of the desks. But our generated scenes rarely place a chair behind a desk. In addition, we find that some of the scenes have chairs on the sides of the desks. This knowledge is inferred from the placement of object-object pairs across all scenes in a room category.

\end{document}